\documentclass[12pt]{iopart}
\usepackage{graphicx}
\usepackage{subfigure}
\usepackage{multirow}
\usepackage{color}
\definecolor{purple}{rgb}{0.4, 0.0, 0.4}
\begin{document}

\title[Pattern recognition of $^{136}$Xe  double beta decay events]{Pattern recognition of $^{136}$Xe  double beta decay events and background discrimination in a high pressure Xenon  TPC}

\author{S~Cebri\'an, T~Dafni, H~G\'omez\footnote{Present address: Laboratoire de l'Acc\'el\'erateur Lineaire (LAL). Centre Scientifique d'Orsay. B\^{a}timent 200 - BP 34. 91898 Orsay Cedex, France.}, D~C~Herrera, F~J~Iguaz, I~G~Irastorza, G~Luz\'on, L~Segui\footnote{Present address: Physics Department , University of Oxford, The Denys Wilkinson Building, Keble Road, Oxford, OX1 3RH, UK.
}, A~Tom\'as\footnote{Present address: Blackett Laboratory, Imperial College London, SW7 2AZ, UK.} }

\address{Laboratorio de F\'isica Nuclear y Astropart\'iculas, Departamento de F\'isica Te\'orica,
Universidad de Zaragoza, 50010 Zaragoza, Spain.}

\ead{luzon@unizar.es}
\begin{abstract}
High pressure xenon gas Time Projection Chambers (TPC) for the detection of the neutrinoless double beta decay of $^{136}$Xe provide good energy resolution and detailed topological information of events. The ionization topology of the double beta decay event of $^{136}$Xe in gaseous xenon has a characteristic shape defined by the two straggling electron tracks ending in two larger energy depositions. With a properly pixelized readout, this topological information is invaluable to perform powerful background discrimination. In this study we carry out detailed simulations of the signal topology, as well as of the competing topologies from gamma events that typically compose the background at these energies. We define observables based on graph theory concepts and develop automated discrimination algorithms which reduce the background level in the region of interest in around three orders of magnitude while keeping signal efficiency of 40\%.
 This result supports the competitiveness of current or future $\beta\beta$ experiments based on gas TPCs, like the Neutrino Xenon TPC (NEXT) currently under construction in the Laboratorio Subterr\'aneo de Canfranc (LSC).

\end{abstract}

\submitto{\JPG}
\maketitle

\section{Introduction}\label{intro}
The observation of neutrino oscillations in atmospheric and solar neutrinos implies that they have mass and mix, but it has left open questions about their mass hierarchy models and their nature, Dirac or Majorana. This has given new motivation for the searches of the neutrinoless double beta decay \cite{Elliot:2003sre}--\cite{ Avignone:2008cs}. Double beta decay is a rare transition between two nuclei with the same mass number, $(Z,A)\rightarrow (Z+2,A)+2e^-+2\bar\nu)$, implying the emission of two electrons and two neutrinos ($\beta\beta 2\nu$). This process has been measured for several nuclei with lifetimes of the order of $10^{18}$--$10^{21}$ years. A hypothetical process without neutrinos in the final state ($\beta\beta 0\nu$) violates lepton number and requires, in the simplest theoretical models, the virtual exchange of massive Majorana neutrinos.

Though different experiments have explored the region of effective neutrino mass above 250\,meV, this decay has never been observed. Current lifetime limits for this mass are around $10^{23}$--$10^{25}$ years depending on the isotope. The near--future aim is to be sensitive to neutrino masses down to 50-100\,meV.  To achieve this goal a new generation of experiments is being designed and built with a detection mass around 100\,kg, improved energy resolution, and lower background level. Several approaches have been considered up to now: semiconductor detectors or bolometers with an excellent energy resolution (GERDA \cite{Ackermann:2013} or CUORE \cite{Ardito:2004}) using segmented detectors to reject multiple--hit events, tracking detectors with pattern recognition capabilities to discriminate background events versus signal (SUPERNEMO \cite{Arnold:2010}), liquid scintillator detectors with a $\beta\beta$ decaying isotope disolved in their active volume (SNO+ \cite{Hartnell:2012} and KAMLAND  \cite{Terashima:2008}), or Xenon TPCs with improved resolution and high tracking capabilities (NEXT \cite{Granena:2009fg}) or a possible ion product tagging (EXO \cite{Danilov:2000}). See reference \cite{Avignone:2008cs} for a review of present experiments and an overview of future programs. Moreover, it would be desirable that any of these designs be easily scalable to reach the inverted--hierarchy scale region (about several tons of isotope mass and extra handles to deal with background events).

In this paper we will focus on the high pressure xenon gas option which offers several advantages. On one hand the isotope: xenon has a natural abundance of 8.9\% in $^{136}$Xe ($\beta\beta$ emitter) and can be enriched by  centrifugation at reasonable cost; it has also a high two--electron energy end point \cite{Redshaw:2007mr}($Q_{\beta\beta}=2458$\,keV), and a large $\beta\beta 2\nu$ lifetime ($T^{\beta\beta2\nu}_{1/2}\sim 2.3\times 10^{21}$\,years, as recently measured by EXO \cite{Ackerman:2011na} and KamLAND \cite{Gando:2012ag})  which minimizes the overlap of the populations of the two neutrinos and the neutrinoless modes; it does not have other long--live radioactive isotopes and responds to the passage of particles by a prompt $< 100$\,ns scintillation light, that can be used as start of the event. These features have motivated several experiments to use this double beta emitter as the aforementioned EXO and KamLAND. On the other hand, the use of high pressure xenon gas improves the energy resolution in this medium \cite{Bolotnikov:1997ab} and provides new handles to reduce background thanks to its extraordinary pattern recognition capabilities which allow distinguishing the characteristic signature of a $\beta\beta$ event from other background events. Moreover, a xenon gas TPC offers scalability to large masses of $\beta\beta$ isotope.
The first experiment of this type was the Gothard TPC \cite{Vuilleumier:1993jcv} in the 90's. Currently NEXT is following a modern version of this same detection technique.

NEXT stands for Neutrino Experiment with a Xenon TPC, and its purpose is to build a 100\,kg high pressure xenon gas TPC, enriched with this isotope, to measure both modes of the double-beta decay. In this medium primary electrons produce excitation (emission of $\sim$ 178\,mm VUV primary scintillation light) and ionization electrons which are drifted towards the TPC anode by an electric field entering in a region with a more intense electric field where further VUV photons are generated isotropically by electroluminescence. The conceptual design  proposes the measurement of the energy (electroluminiscent light) and the event start (primary scintillation) in a sparse plane of PMTs (the energy plane) located behind the cathode. The tracking plane will be placed at the anode plane and consists of 1\,mm$^2$ Silicon Photomultipliers forming arrays of 1\,cm pitch. The experiment was proposed to the Laboratorio Subterr\'aneo de Canfranc (LSC), Spain, in 2009 \cite{Granena:2009fg}, with a source mass of the order of 100 kg. Three years of intense R\&D have resulted in a Conceptual Design Report (CDR) \cite{Alvarez:2012va} and a Technical Design Report (TDR) \cite{Ferreira:2012va}.

\begin{figure}[htb!]
\centering
\includegraphics[width=75mm, angle=270]{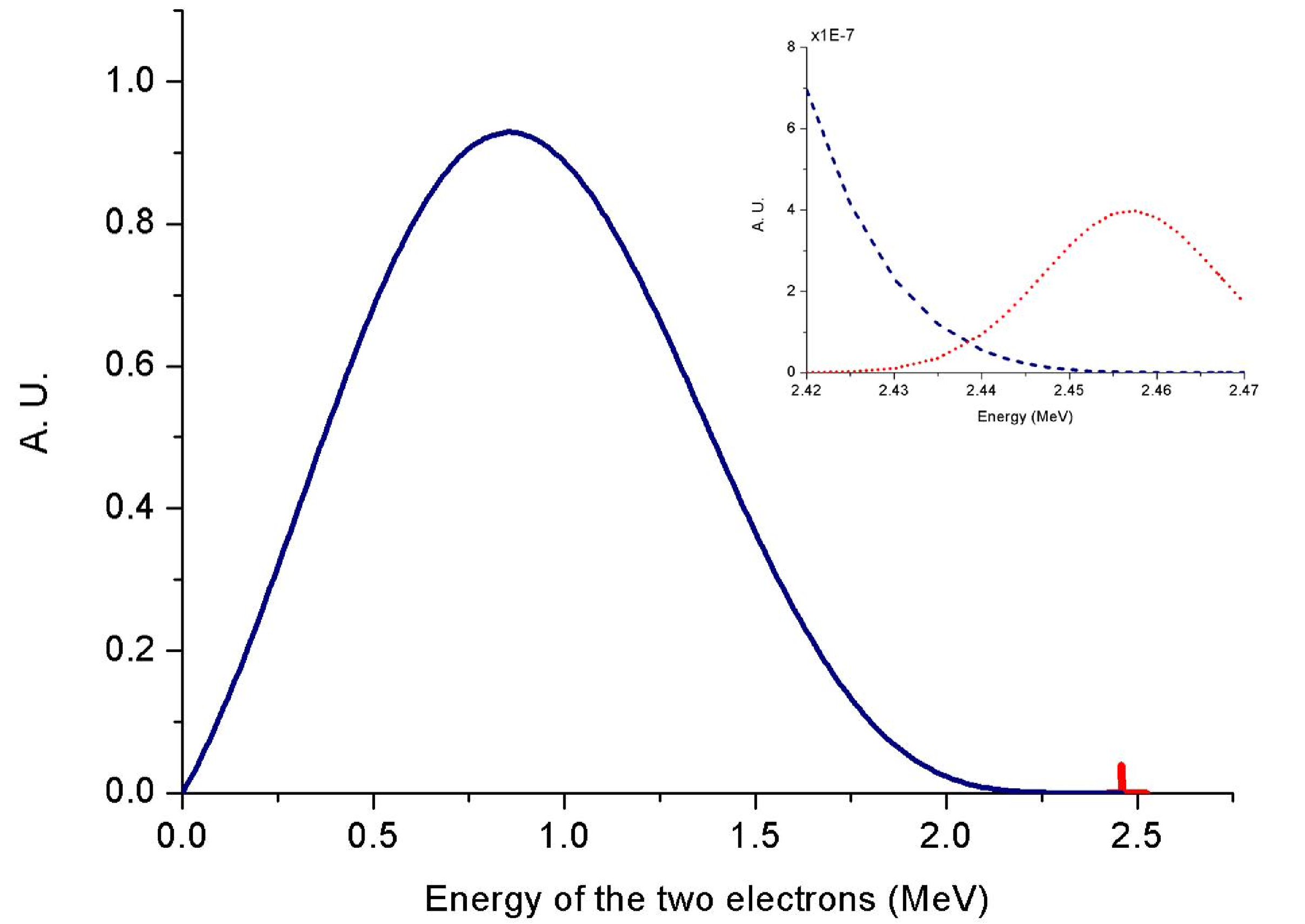}
\caption{Two electrons energy spectrum for  $^{136}$Xe $\beta\beta2\nu$ decay with end point $Q_{\beta\beta}=2458$\,keV; the $\beta\beta 0 \nu$ peak is shown (not to scale). In the upper right inset the region around $Q_{\beta\beta}$ is zoomed in, showing the overlap of both populations due to finite detector energy resolution.  }
\label{figure1}
\end{figure}

Though the $\beta\beta 0\nu$ signal can be well characterized as a peak at the end of the $\beta\beta 2\nu$ spectrum (figure \ref{figure1}), its extremely low rate makes its detection a big experimental challenge: an effective neutrino mass of near $m_{\beta\beta}=100$\,meV implies just a few counts per year for an isotope mass of 100\,kg; and around a 1\,ton of isotope is needed to get a few counts per year assuming an effective neutrino mass of 50\,meV. Due to finite energy resolution, events from the $\beta\beta2\nu$ distribution tail may constitute an irreducible background for the detection of the $\beta\beta 0\nu$ signal events. The need to keep this background well below the stated signal levels sets a maximum accepted energy resolution of $\sim$4.5\% FWHM for a 100\,kg experiment, and of $\sim$2.5\% for a one ton experiment \cite{Avignone:2008cs}, a requirement certainly at hand for the Xe gas TPC technique. In general, a background level below $10^{-3}$\,c\,keV$^{-1}$\,kg$^{-1}$\,yr$^{-1}$ at the $Q_{\beta\beta}$  is required for a 100\,kg experiment to achieve sensitivity to $m_{\beta\beta}\sim 100$\,meV in a few years data taking campaign. For a detector of a few tons to explore the inverse hierarchy region a background level down to $\sim10^{-5}$\,c\,keV$^{-1}$\,kg$^{-1}$\,yr$^{-1}$ could be needed. In addition to standard background reduction strategies, like operation underground, use of shielding and selection of radiopure detector components through screening, pattern recognition techniques to discriminate signal from background events must be exploited at the maximum.

 Pattern recognition for $\beta\beta$  searches has been widely studied, e.g. for germanium experiments \cite{Gomez:2007hg, Agostini:2011ma}, using the event pulse shape to distinguish single--site from multi-site events. In gaseous xenon, the Gothard group \cite{Wong:1993htw} pioneered the study and use of event topology, and introduced some key ideas like the presence/absence of two higher energy depositions at the ends of the $\beta\beta$  electron tracks. They successfully demonstrated the application of these concepts to real data (in a 5.3 kg gas Xe TPC), although the discrimination was done through visual inspection. The aim of the work presented in this paper is to supersede the topological recognition techniques initiated by the Gothard group, introducing new ideas for reconstruction and analysis, including the full simulation of the expected signals and background in a high pressure Xenon TPC, and elaborate automated discrimination algorithms with this topological information \cite{Iguaz}\cite{Segui}.

In section \ref{sec:physics}, the predicted $^{136}$Xe $\beta\beta0\nu$ signal and other expected background events in the energy Range of Interest (RoI) are described stressing their main distinguishing topological features. In section \ref{sec:Discrimination}, the algorithms used to characterize the events are described, and the selection criteria used to discriminate signals from background events defined. Their rejection power factor is presented in section \ref{sec:results} together with the signal efficiency comparing both high and low diffusion media. Finally some conclusions and an outlook of possible improvements in the analysis are gathered.

\section{Signal and background topological signature in detectors equipped with pixelized readouts} \label{sec:physics}
The neutrinoless double beta decay of $^{136}$Xe entails the emission of two electrons from a common vertex, sharing a total energy of 2458\,keV \cite{Redshaw:2007mr}. During their travel through the high pressure gas they typically produce a characteristic ionization pattern that, although with variations from event to event, allows to define a prototype topology on which to base our identification criteria. The prototype topology for the signal events is a continuous straggling ionization track of around 30\,cm (for a gas pressure of 10\,atm) and, at both ends of the track, high energy depositions (blobs) due to the Bragg peak of the electrons. Obviously, also the $\beta\beta 2\nu$ events have the same topology; though, as mentioned before, these events will not constitute a relevant background in the RoI provided the detector has sufficiently good energy resolution. Any other background event near the $Q_{\beta\beta}$ energy will consist of one or more single--electron tracks, and their prototype topology will be that of a single ionization track with only one higher energy blob at one of the track's ends. These two conceptual topologies offer a clear distinction between signal and background events and are the basis for the design of the discrimination criteria. However, the distinction is in practice limited by additional physical interactions (secondary radiation, multi-site interactions,...) or detector effects (diffusion in the gas and readout digitization). Whether the discrimination criteria can be designed to be inmune to these effects, and to which extent, is the main goal of our study.

In the following the signal and background events able to deposit their energy in a
energy RoI around $Q_{\beta\beta}$ will be simulated and studied in detail. All our results are expressed as fractions of the events inside this RoI that are rejected or accepted by the different criteria. The width of this RoI should include most signal events and therefore it is linked to the energy resolution of the experiment. Due to the fact that the background population is either flat in energy around $Q_{\beta\beta}$, or, in the case of Bi, in a peak too close to the $\beta\beta0\nu$ peak to be resolved by energy resolution, our results are, in first approximation, independent on the choice of the RoI. For this work we have taken the practical choice of 2400-2500\,keV for the RoI, which corresponds to a $\pm$2\% around $Q_{\beta\beta}$ and should generously encompass the $\beta\beta0\nu$ peak for the target energy resolution of modern Xe gas TPCs: NEXT-100 targets a resolution of 0.5\% FWHM, while some of its prototypes are already showing an energy resolution below the 1\% FWHM \cite{Ferreira:2012va}\footnote{Other experiments using xenon show a more modest energy resolution, as KAMLAND-Zen, where the $^{136}$Xe isotope has been disolved in liquid scintillator, with a 9.9\% FWHM resolution at the $Q_{\beta\beta}$ of $^{136}$Xe \cite{Gando:2012ag}, or EXO-200, based on a liquid Xe TPC, which achieves currently an energy resolution of 4\% FWHM \cite{Ackerman:2011na} (an energy resolution as low as 3.3\% FWHM could be reached \cite{Conti:2003}).}. Another choice in our analysis is the gas pressure, assumed to be 10\,bar. This is a reasonable design choice (followed by the NEXT TPC) out of a compromise between detector compactness and quality of the topological information. However a detailed study of the topological figure of merit versus gas pressure is still pending.

\subsection{Expected background events}

Any emission of particles with energies above the $Q_{\beta\beta}$ constitutes potentially a source of background for the experiment. We are concerned in particular with those able to leave single ionization tracks of energies in the RoI (electrons or gammas) in the fiducial volume of the detector, as they may end up mimicking the two--blob signal topology by means of secondary photon emission or straggling. In the following points we discuss the different background populations to identify the relevant ones and focus our study on them:

\begin{enumerate}
\item\emph{Radioactive contamination of laboratory walls.}

In this case, only photons are able to reach the detector. The contribution of the external gamma flux can be much more important than internal contaminations if the shielding is not thick enough. As an example, in the Underground Laboratory of Canfranc, the gamma flux of 0.17 $\gamma$ cm$^{-2}$ \cite{Luzon:2006gl} produces a background level around $ 10^4$\,c\,keV$^{-1}$\,kg$^{-1}$\,yr$^{-1}$ in the RoI in absence of shielding. This value is several orders of magnitude higher than the total internal contribution but it can be reduced to less than $ 10^{-4}$\,c\,keV$^{-1}$\,kg$^{-1}$\,yr$^{-1}$ with a lead shielding of at least 25\,cm thickness \cite{Ferreira:2012va}. Also airborne $^{222}$Rn, which decays into $^{214}$Bi, can increase the background level since it can reach the detector vessel. A continuous flushing of liquid nitrogen inside a sealed plastic bag surrounding the detector prevents the radon intrusion. Thus only internal contaminations will be included in our analysis.

\item\emph{Radioactive contamination of shielding and detector materials.}

The materials used in the detector setup and shielding contain impurities of radioactive elements, mainly the natural decay chains of $^{232}$Th and $^{238}$U. They decay emitting alpha or beta particles, as well as photons due to subsequent decays to ground levels. Both alpha particles and electrons are quickly absorbed in media and only those emitted near surfaces facing the sensitive volume are detected, but easily rejected by fiducialization. Then, only a few energetic enough gamma emissions are relevant: the 2614.5\,keV photon emitted in the 99\% of the $^{208}$Tl $\beta$ decays; and the 2447.8\,keV (1.57\% of intensity) gamma line following a low energy $^{214}$Bi $\beta$ decay. Also, for this last isotope,  some energetic $\beta$ emissions with $Q_{max}^\beta$ of 3272\,keV (18.2\%), 1894.3\,keV (7.43\%) or 2662.7\,keV (1.7\%) produce electrons inside the RoI. Other contaminants, like $^{60}$Co, can be neglected, according to preliminary simulations, since its two gamma emission (1173\,keV and 1332\,keV), when both gammas are interacting  in the gas, deposits up to 2505\,keV (slightly above the RoI) and produces at least two well separated energy depositions in the gas even in the case of high diffusion media. The rejection factor in that case is 2 orders of magnitude higher than that of the $^{214}$Bi (limited by simulation statistics).

$^{222}$Rn emanated by materials inside the vessel can also constitute a relevant background source. It can diffuse in the detector gas and reach the sensitive volume. Its $\alpha$ decay is harmless as it produces a very distinctive almost point--like signal (less than 5\,mm of range in xenon at 10\,atm) of energy much above the RoI, but the resulting positively charged $^{218}$Po ions can drift and deposit on surfaces (especially on the TPC cathode). Among the following progeny is $^{214}$Bi. Therefore, for the purpose of this study, this background source is equivalent to a surface $^{214}$Bi contamination.

\item  \emph{High energy photons due to underground muon interactions.}

Cosmic muons passing through matter produce high energy photons which result from muon bremsstrahlung, direct pair production by muons and muon--nuclear interaction. At Canfranc Laboratory underground muons present a mean energy of 290\,GeV and a flux intensity of around $5\times10^{-7}$\,cm$^{-2}$s$^{-1}$\,sr$^{-1}$. Preliminary simulations show that their contribution to background is low (of the order of $10^{-4}$\,c\,keV$^{-1}$\,kg$^{-1}$\,yr$^{-1}$). Moreover the use of active veto detectors around the experiment would veto not only muons but also any event induced by muons in shielding and detector materials. Therefore, this contribution will be neglected in the following.

\item\emph{Neutron activation.}
There are no worrisome cosmogenic isotopes to be activated in the gas and in the detector materials by sea level neutrons. Even cosmoisotopes as $^{60}$Co, very disturbing for some of the $\beta\beta$ experiments, can be neglected in our case. Once materials and gases are stored underground, the radioactive capture by $^{136}$Xe of medium energy neutrons may produce $^{137}$Xe, which decays with a half life of 3.8\,min, 67\% of the times as a beta decay with $Q_\beta=4173$\,keV and a 30\% as a beta decay with $Q_\beta=3717.5$\,keV and a gamma of 455\,keV. The production rate of this isotope in the gas of the TPC is computed to be only 0.6\,nuclei year$^{-1}$ kg$^{-1}$, and makes this contribution irrelevant.

\end{enumerate}
Summarizing the analysis of the different sources, only $^{208}$Tl and $^{214}$Bi, are relevant as possible background contributions in the experiment. For this reason, the studies which follow will be focused on these isotopes.

\subsection{Simulation overview}\label{sec:sim}

\begin{figure}[tb]
\centering
\includegraphics[width=80mm]{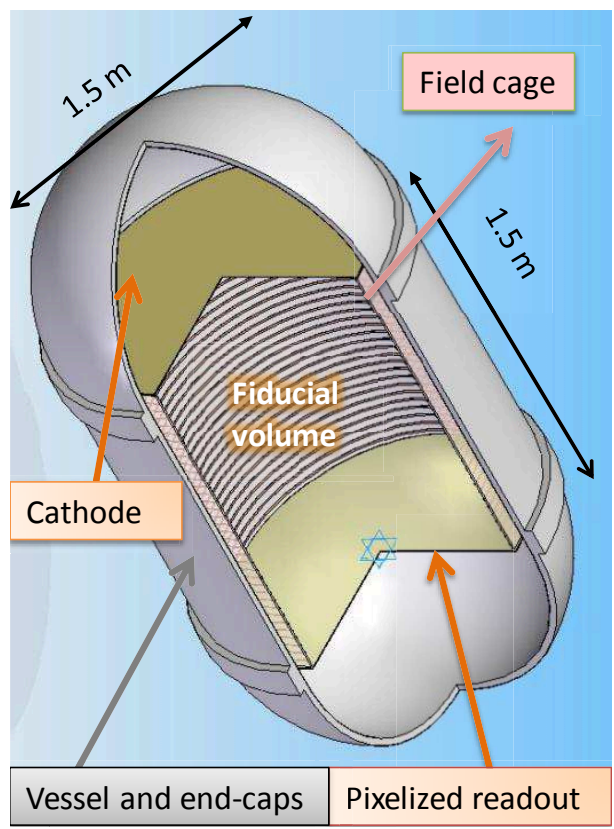}
\caption{A view of the geometry simulated in GEANT4, explained in detailed in the text. The different volumes of the TPC can be observed: the fiducial volume, the field cage, the cylindrical TPC with its two endcaps, the cathode and the readout.}
\label{figure2}
\end{figure}

The aim of this work is to study generic properties of event topologies and discrimination criteria, as independent as possible from the specific detector design choices. As some of the aspects of the work may depend on geometrical issues (e.g., the origin of the $^{214}$Bi or $^{208}$Tl contaminations or the size of the fiducial volume), to understand to some extent these dependencies, we have implemented a simplified generic geometry of a gaseous TPC in our simulations, that is shown in figure \ref{figure2}. We want to stress that the aim of this work is not to elaborate a background model, for which a more detailed geometry and a complete material account would be needed.

The main element of the simulated geometry is a cylindrical vessel, closed by two semispherical end caps, as shown in figure \ref{figure2}. The vessel walls are made of copper of 3\,cm thickness. The chosen dimensions of the cylinder are a length of 1.5\,m and a diameter of 1.6\,m. Inside the TPC, there is a field cage made of teflon, with copper rings embedded on it, to shape the drift field along the $z$--direction. At both sides of the field cage there are two surfaces representing the cathode and the pixelized readout (1$\times$1\,cm$^2$  area) placed at the anode. All the space between the walls and the field cage is filled with xenon gas at 10\,bar. The sensitive volume, where the interactions of particles are recorded, is the cylindrical volume inside the field cage, whose dimensions are 1.5 m of drift length and a diameter of 1.38\,m. This volume of gas corresponds to a total mass of 124\,kg of xenon at 10\,bar.

\begin{figure}[htb!]
\centering
\includegraphics[angle=270, width=100mm]{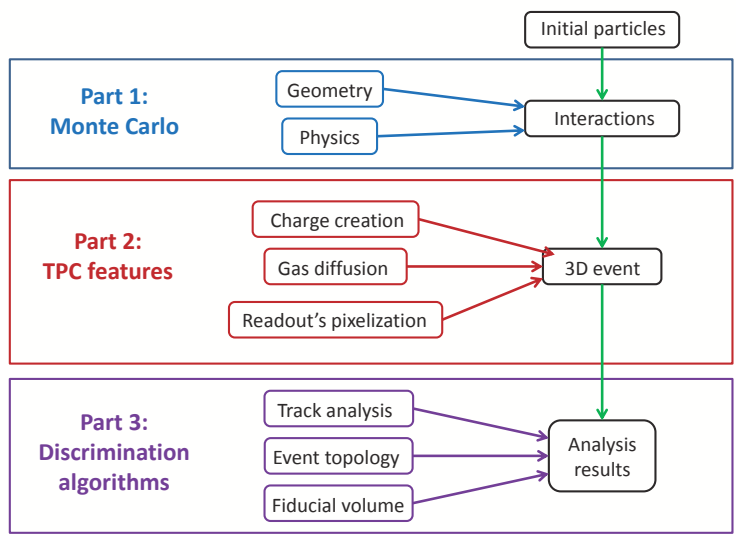}
\caption{Data flow of the simulation and analysis organized into three logical blocks.}
\label{figure3}
\end{figure}

  The simulation and analysis can be divided into three logical blocks: 1) the interaction of ionizing particles with the gas; 2) the simulation of the detector response; 3) the analysis of 3D events. Figure \ref{figure3} illustrates the implementation diagram of these logical blocks in the software.

The first block consists of a Monte Carlo simulation, including all the physical processes involved in the passage of particles through matter. It provides the interaction points and the corresponding energy losses within the gas. This part of the simulation is based on the GEANT4 simulation package \cite{Agostinelli:2003sa,Allison:2006ja} and includes the conceptual geometry described above. The program created has a set of routines for generating vertexes at the different volumes of the geometry and for setting as initial particles those previously defined by the event generator Decay4 \cite{Pokratenko:2000op}. This code generates events from the double beta decay and from the decay of radioactive nuclide of all unstable isotopes.

For the second and third blocks, a software structure based on C++ and ROOT \cite{Root:1997} classes has been used to reconstruct events and to analyze them. This software has been developed by our group as part of a more general framework of software for analysis and reconstruction of events in rare event experiments (RESTSoft). This code is ready to be used in real data, not only on those coming from simulations.

The second block describes the generation of ionization electrons in the gas, the diffusion effects during the drift towards the readout plane and the digitization of the 3D event according to the readout pattern. More specifically, the charge of an energy deposit is generated using the W--value, the mean energy needed to create an ion--electron pair in the gas, extracted from \cite{Christophorou}. The statistical Poissonian fluctuation of the number of generated electron--ion pairs have been taken into account. Then, the electron cloud is Gaussian spread in longitudinal and transversal directions to include charge diffusion effects. The coefficients calculated with Magboltz \cite{SBiagi}, depending on the type of gas, have been used. Finally, the pixelization of the readout has been consider to group charges in 3D pixels.

The last block, including analysis classes developed to recognize the $\beta\beta$ signal and distinguish it from the different background events, is described in detail in section \ref{sec:Discrimination}.

Each of the two kind of contaminations studied ($^{208}$Tl and $^{214}$Bi) are separately simulated from different representative volumes of the geometry, in order to analyse the possible dependence of the algorithm on the origin of the contamination. Four such volumes are considered: the lateral part of the vessel (a volume far from the gas active volume), the field cage (volume facing the gas), the cathode (surface on top of the gas) and the readout (surface at the bottom).
Signal events have been simulated homogeneously in the gas volume. The number of events simulated is fixed differently for each of the simulations so that the initial number of events falling in the RoI is $10^5$, so enough statistics on which to apply the discrimination criteria is gathered.

An important aspect affecting the topology is the diffusion of the electrons drifting in the TPC. It is known that diffusion in pure Xe is relatively large, e.g., the longitudinal and transversal diffusion coefficients respectively are $\sim 300$ and $\mathrm{1000 \mu m/ \sqrt{cm}}$ for a drift field 1\,kV\,cm$^{-1}$\,bar$^{-1}$. An additional low diffusion scenario will be studied, to assess the impact of diffusion in the discrimination capabilities. The use of a small quantity of additives to the Xe, like CF$_4$ or trimethylamine (TMA), might reduce the diffusion coefficients down to $\mathrm{\sim100 \mu m/ \sqrt{cm}}$ in the same conditions of drift field and pressure.

\subsection{Topological characterization of signal and background events}\label{characterization}

\begin{figure}[bt]
        \centering
        \begin{subfigure}[]{
                \includegraphics[height=60mm ,width=75mm]{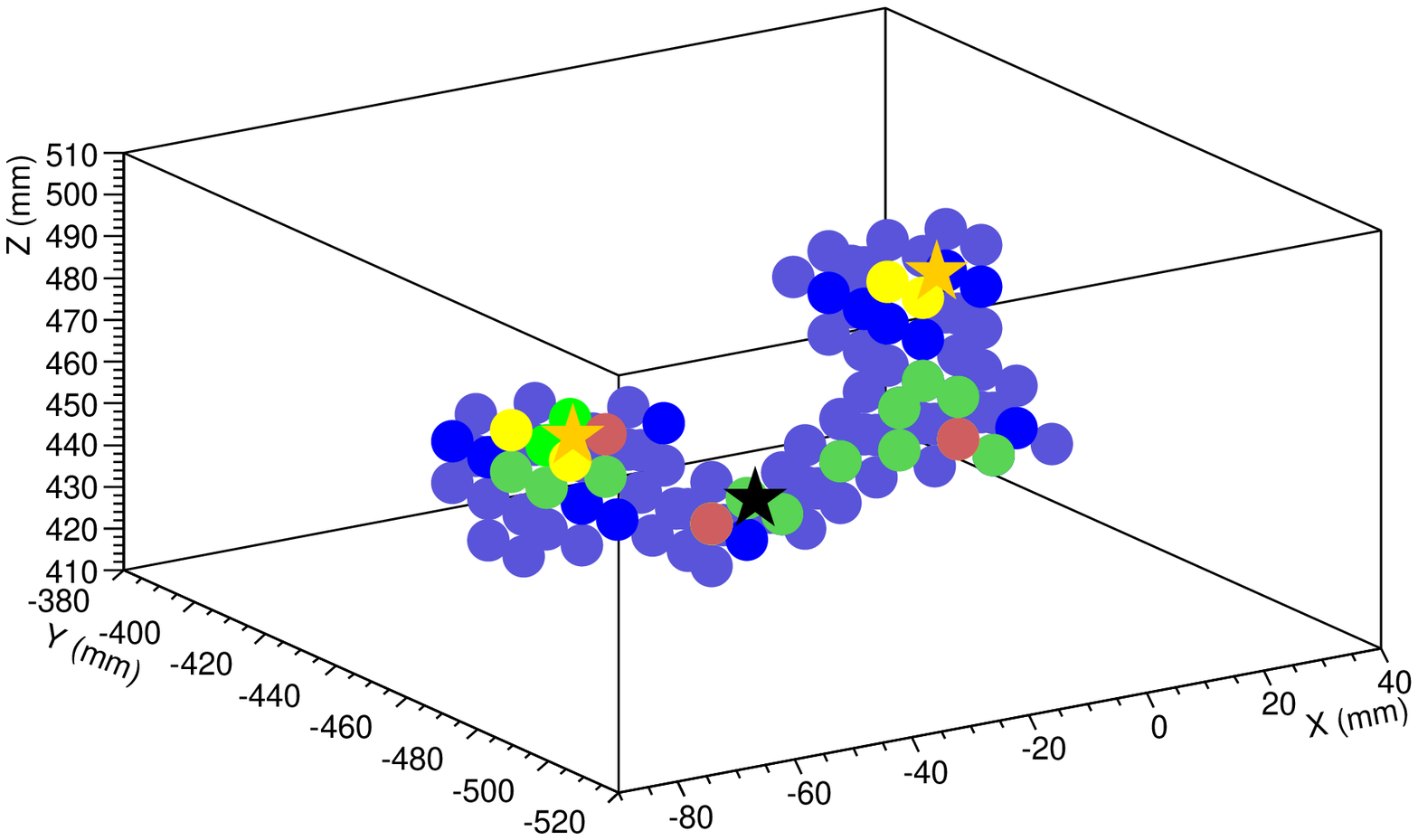}}
        \end{subfigure}
        \begin{subfigure}[]{
                \includegraphics[height=60mm, width=75mm]{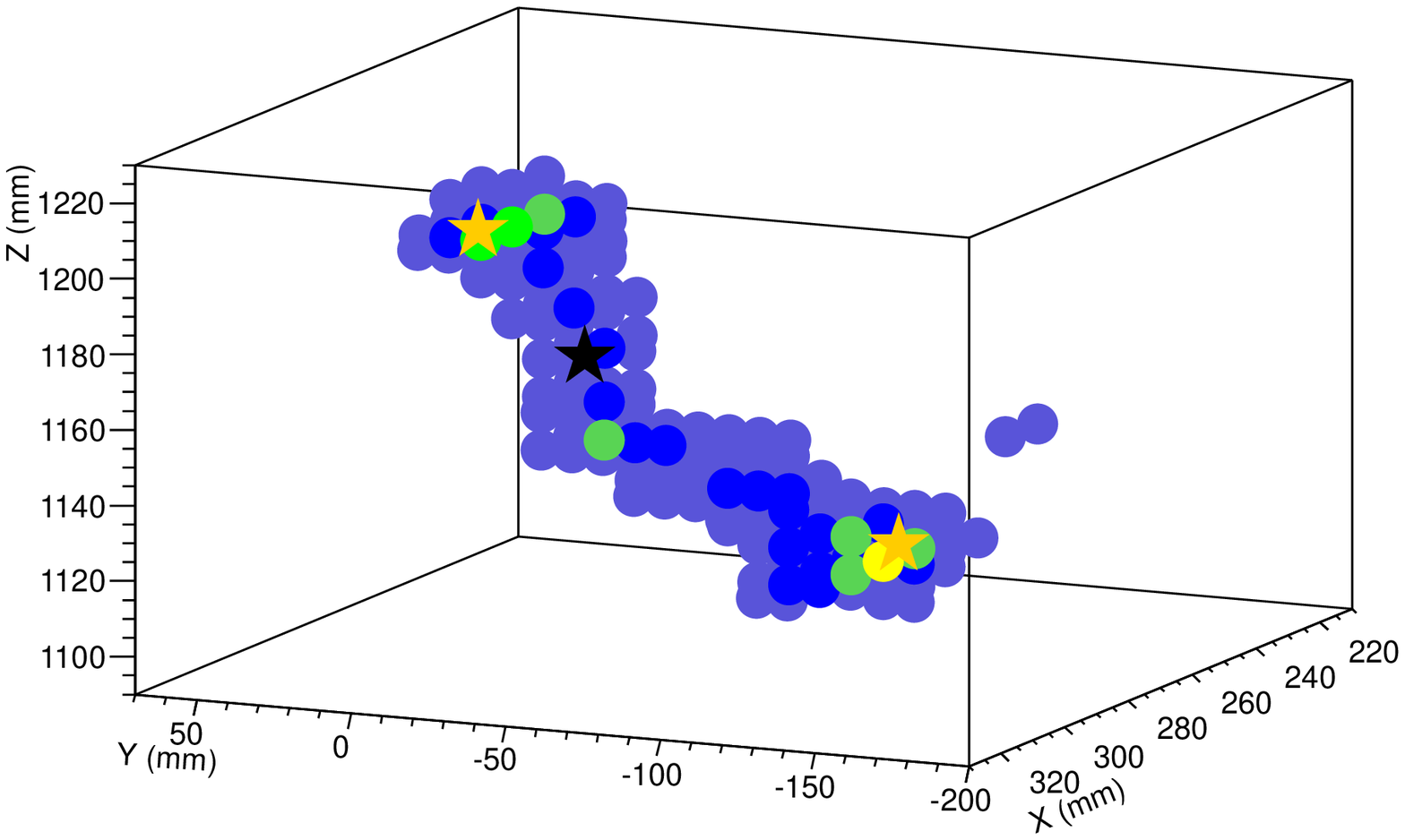}
                \label{figure4b}}
        \end{subfigure}
        \
        \begin{subfigure}[]{
                \includegraphics[height=60mm,width=65mm]{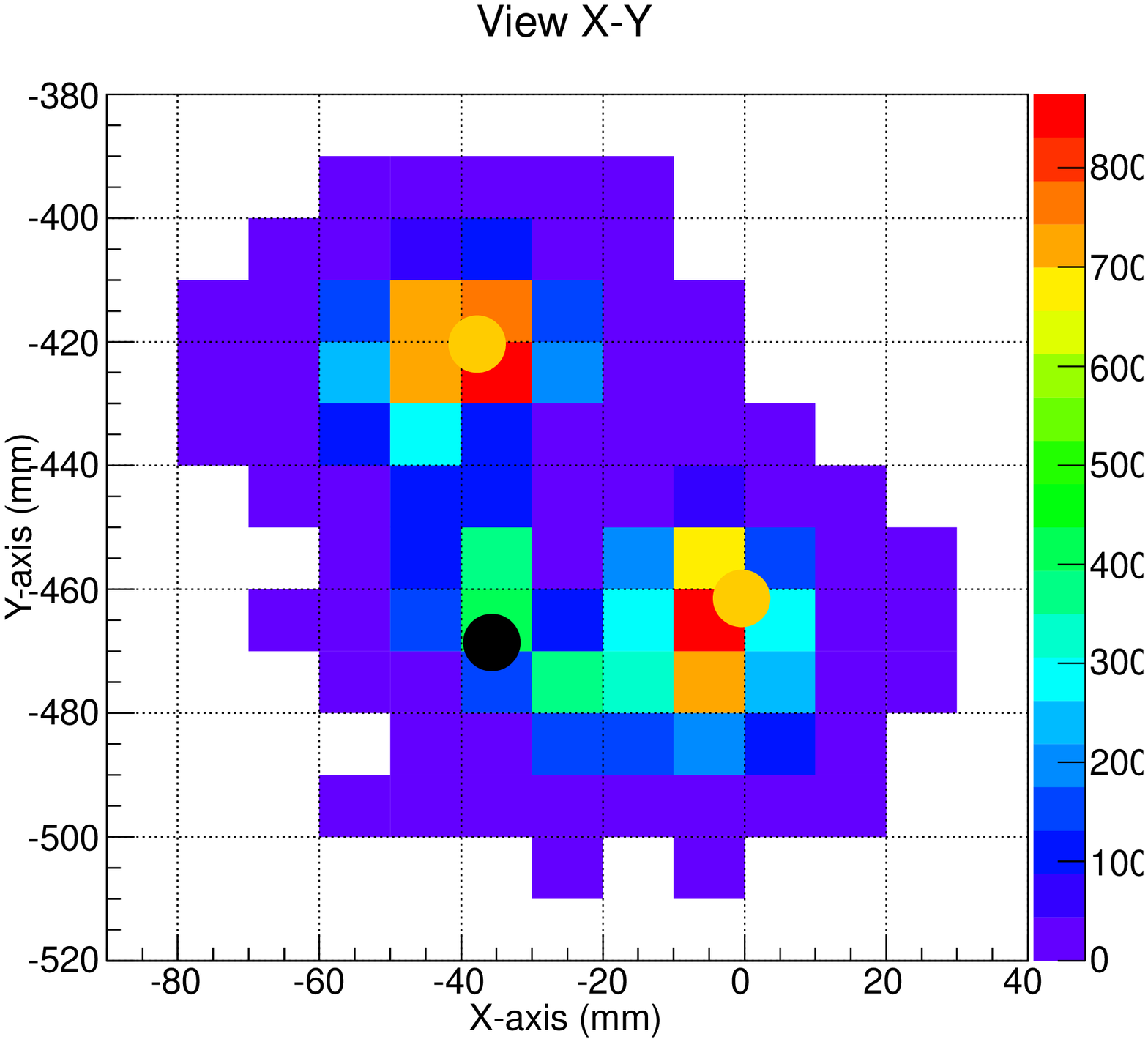}
                \label{figure4c}}
        \end{subfigure}
        \begin{subfigure}[]{
                \includegraphics[height=60mm, width=80mm]{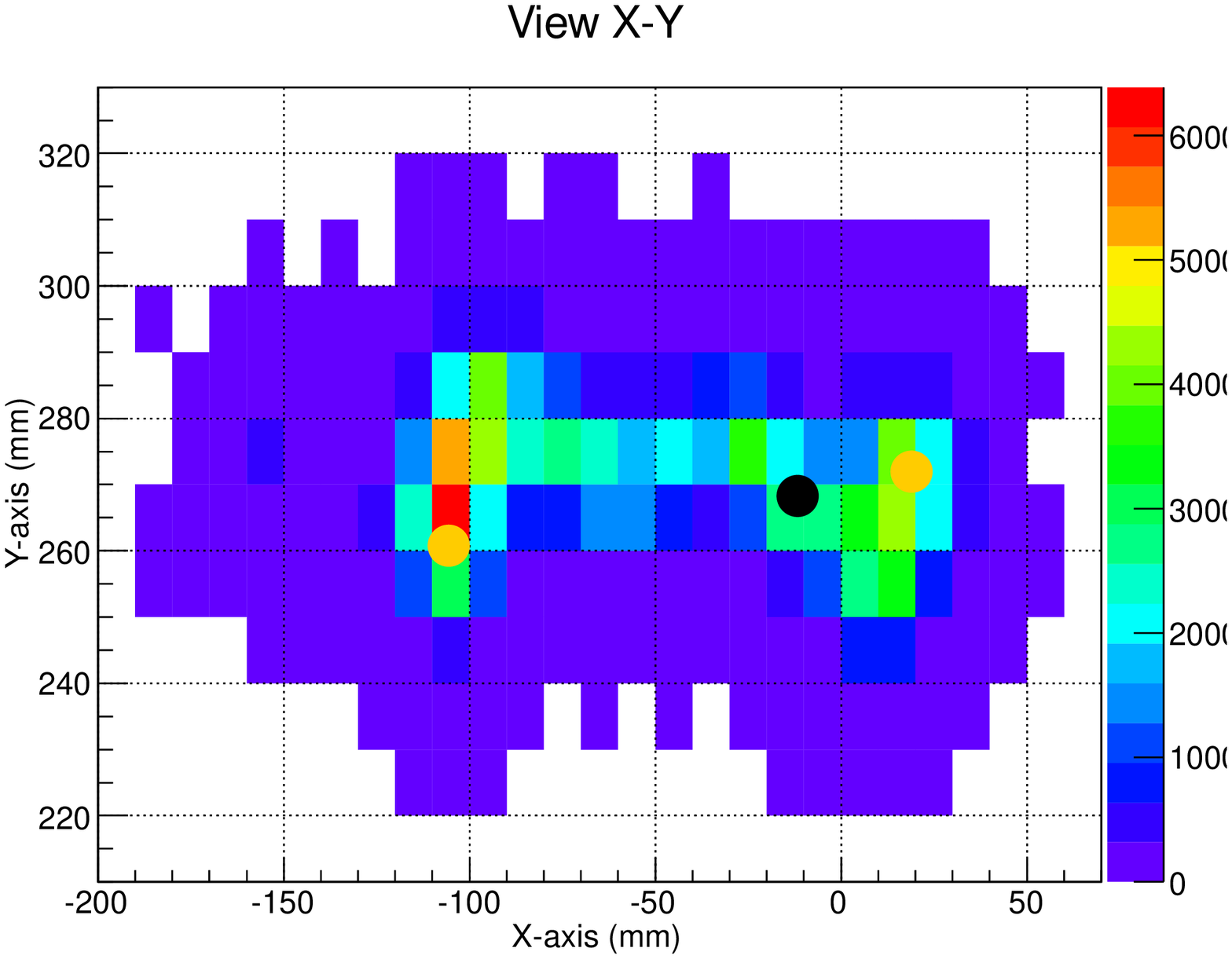}
                \label{figure4d}}
        \end{subfigure}
        \caption{Three--dimensional representation (top) and two--dimensional projection (bottom) of $^{136}$Xe $\beta\beta0\nu$ events simulated in a 10\,bar Xenon TPC. On the left: two electrons of similar energy has been emitted from a vertex (black mark) and show big energy depositions at the end of their tracks (yellow marks). On the right, a smaller track far from the main one can also be seen. }\label{figure4}
\end{figure}

A substantial fraction of signal events may differ from the prototype single--track--with--two--blob pattern described before for the $^{136}$Xe $\beta\beta$ signal (see figure \ref{figure4}). More than 80\% of $\beta\beta0\nu$  events in the RoI have secondary photon emission (x-rays or bremsstrahlung photons). Although most of them do not have enough energy to travel far from the main electron track, in some cases they may leave the sensitive volume (reducing the signal efficiency) or cause the event to show more than one disconnected track. In overall, for the geometry considered, only about 60\% of the signal events are classified as single track events. An effective way to protect the algorithm from this feature is to consider as acceptable signal topologies those with a second short track (i.e. with energy lower than 100\,keV) disconnected from the main one, as will be seen in subsection \ref{subsec:Discrimination_tracks}. When this is done, the signal acceptance is increased by 20\% while the impact on the rejection of background events, normally showing more than two \emph{long} tracks, is minimal.

\begin{figure}[bt]
        \centering
        \begin{subfigure}[]{
                \includegraphics[height=55mm,width=75mm]{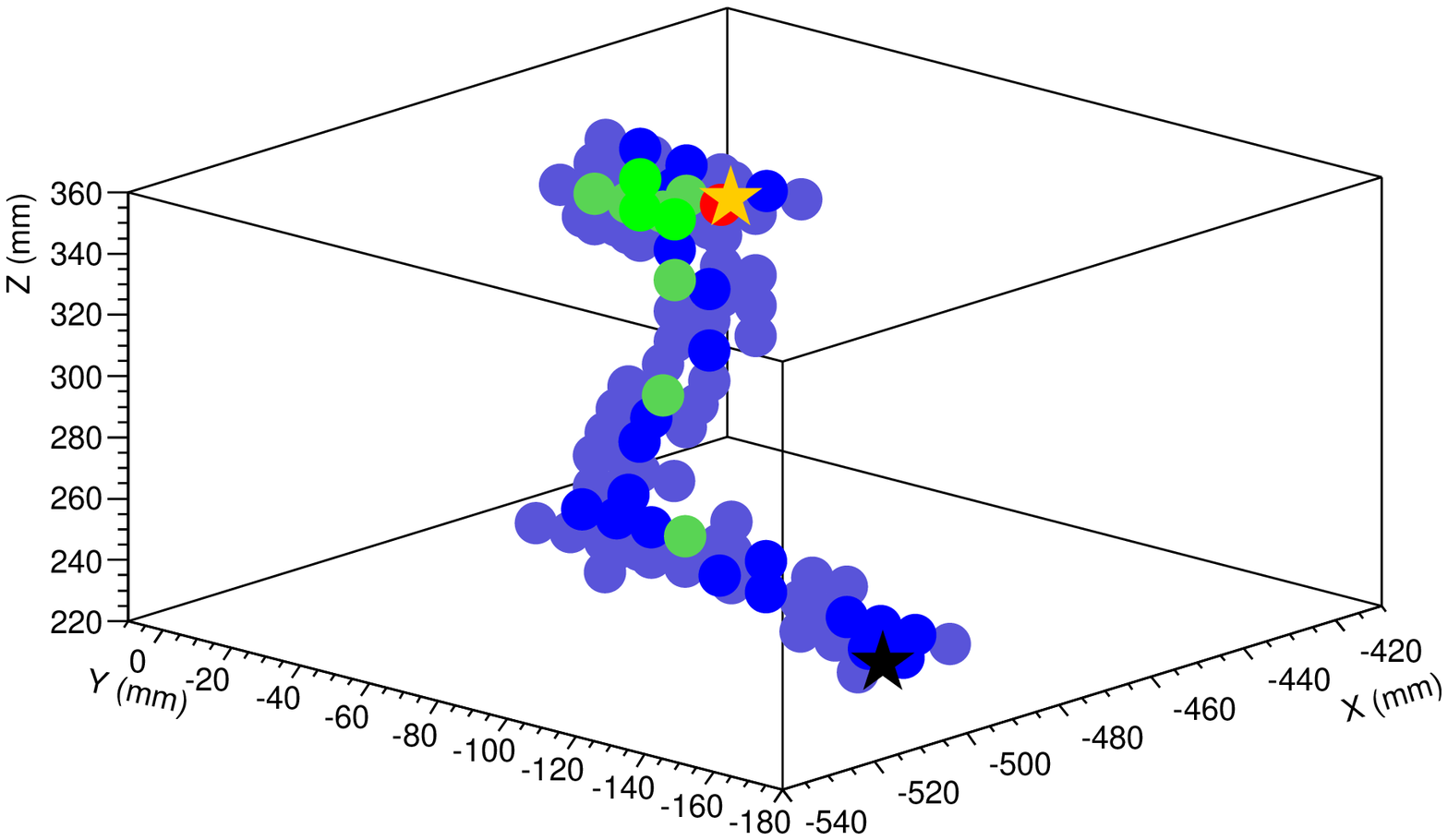}
                \label{figure5a}}
        \end{subfigure}
        \begin{subfigure}[]{
                \includegraphics[height=50mm,width=70mm]{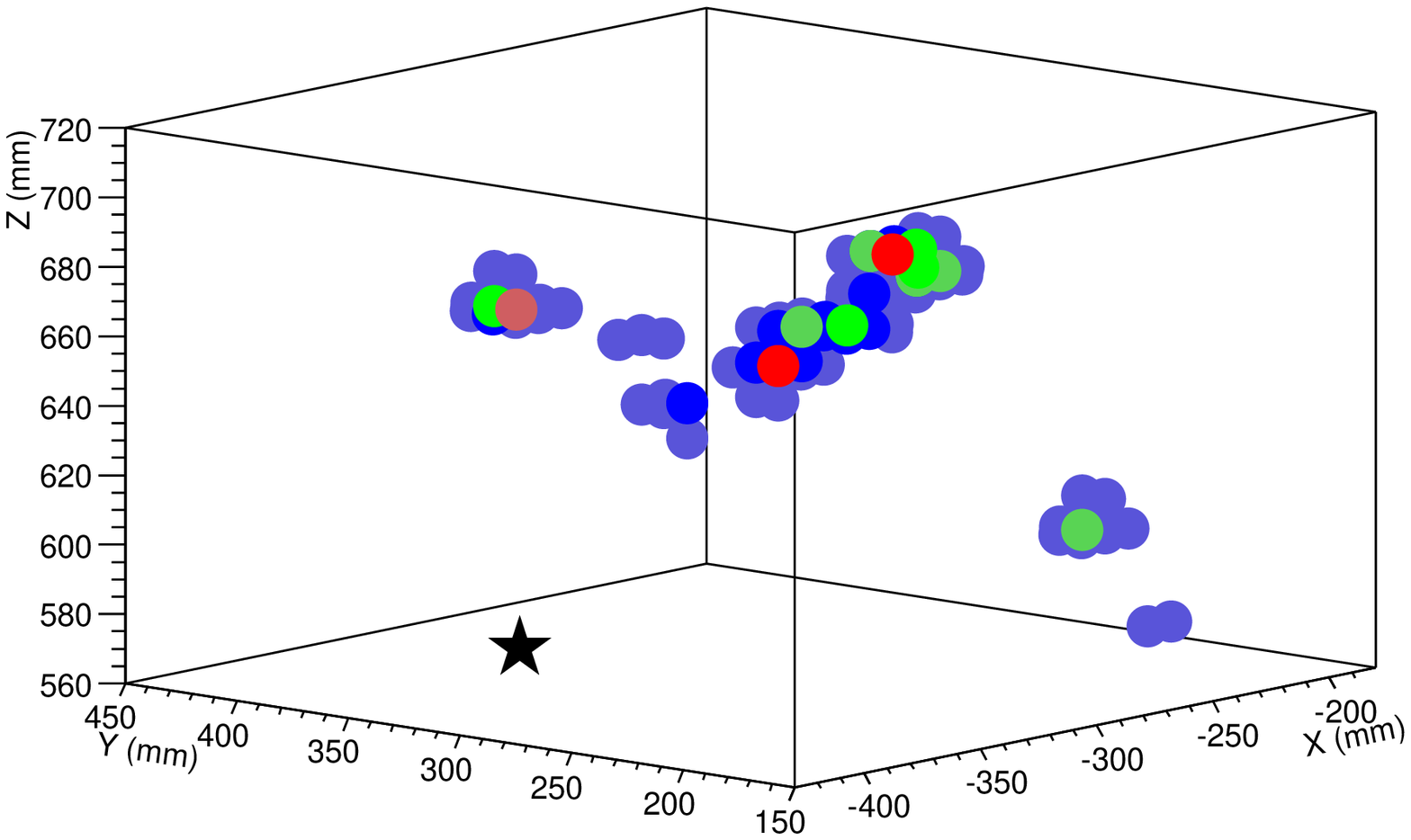}
                \label{figure5b}}
        \end{subfigure}
        \
        \begin{subfigure}[]{
                \includegraphics[height=60mm,width=70mm]{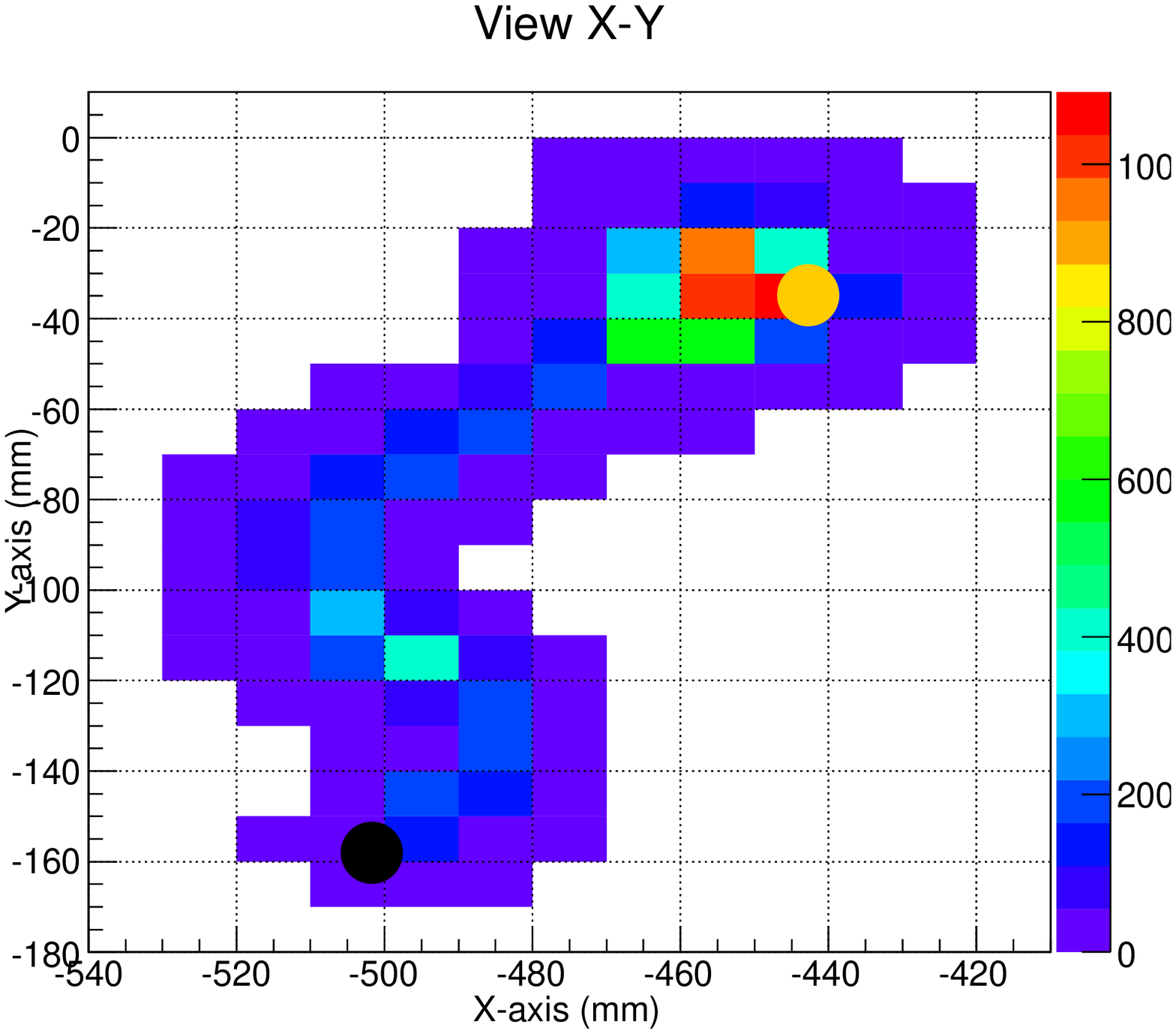}
                \label{figure5c}}
        \end{subfigure}
        \begin{subfigure}[]{
                \includegraphics[height=60mm,width=75mm]{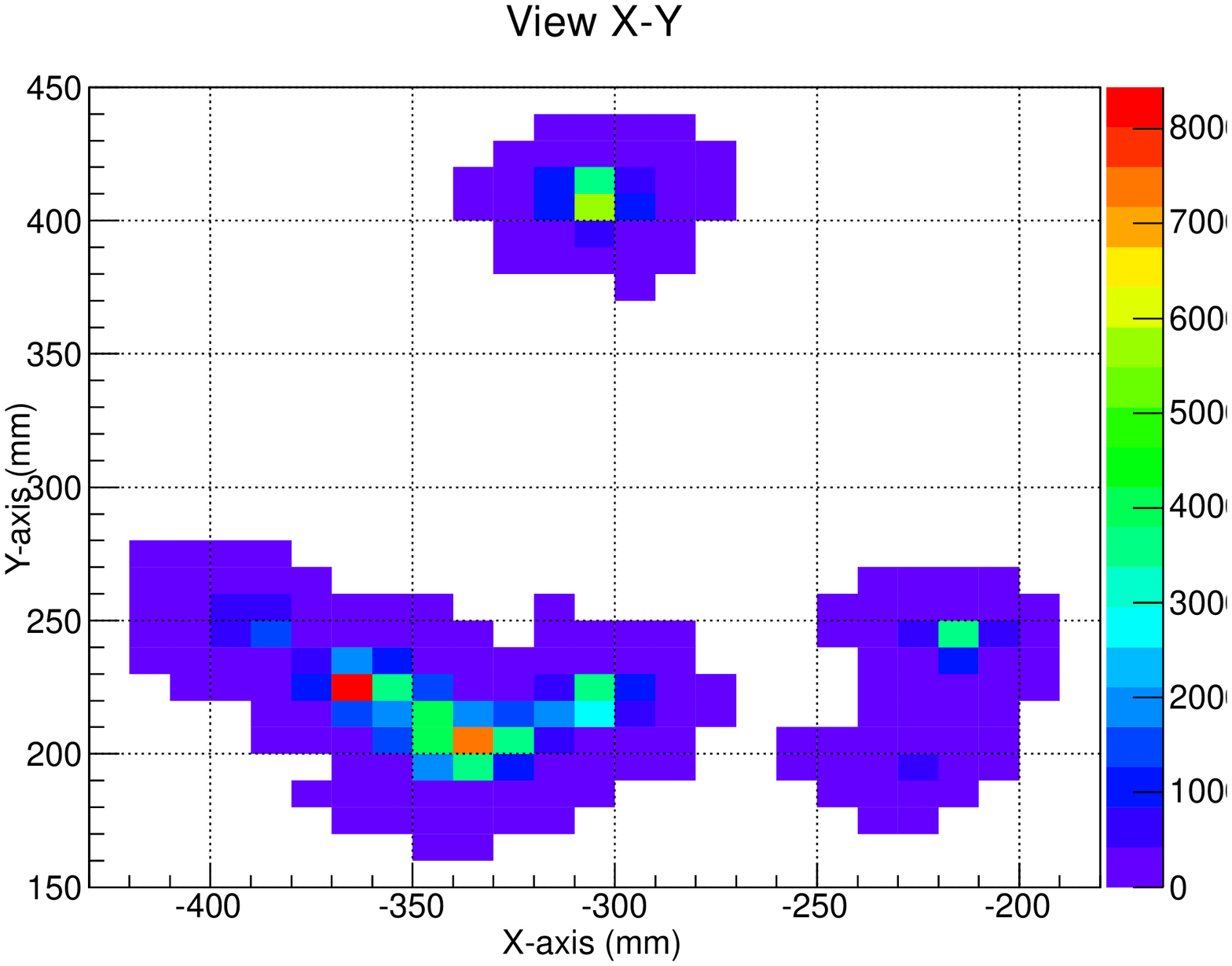}
                \label{figure5d}}
        \end{subfigure}
        \caption{Three--dimensional representation (top) and two--dimensional projection (bottom) of simulated background events in a 10\,bar Xenon TPC. On the left, a $^{214}$Bi electron coming out from the lateral wall leaves a continuous track and shows a higher deposition at the end. On the right, an event caused by a $^{208}$Tl 2614.5\,keV photon presents several tracks. }\label{figure5}
\end{figure}

\begin{figure}[bt]
        \centering
        \begin{subfigure}[]{
                \includegraphics[height=50mm,width=70mm]{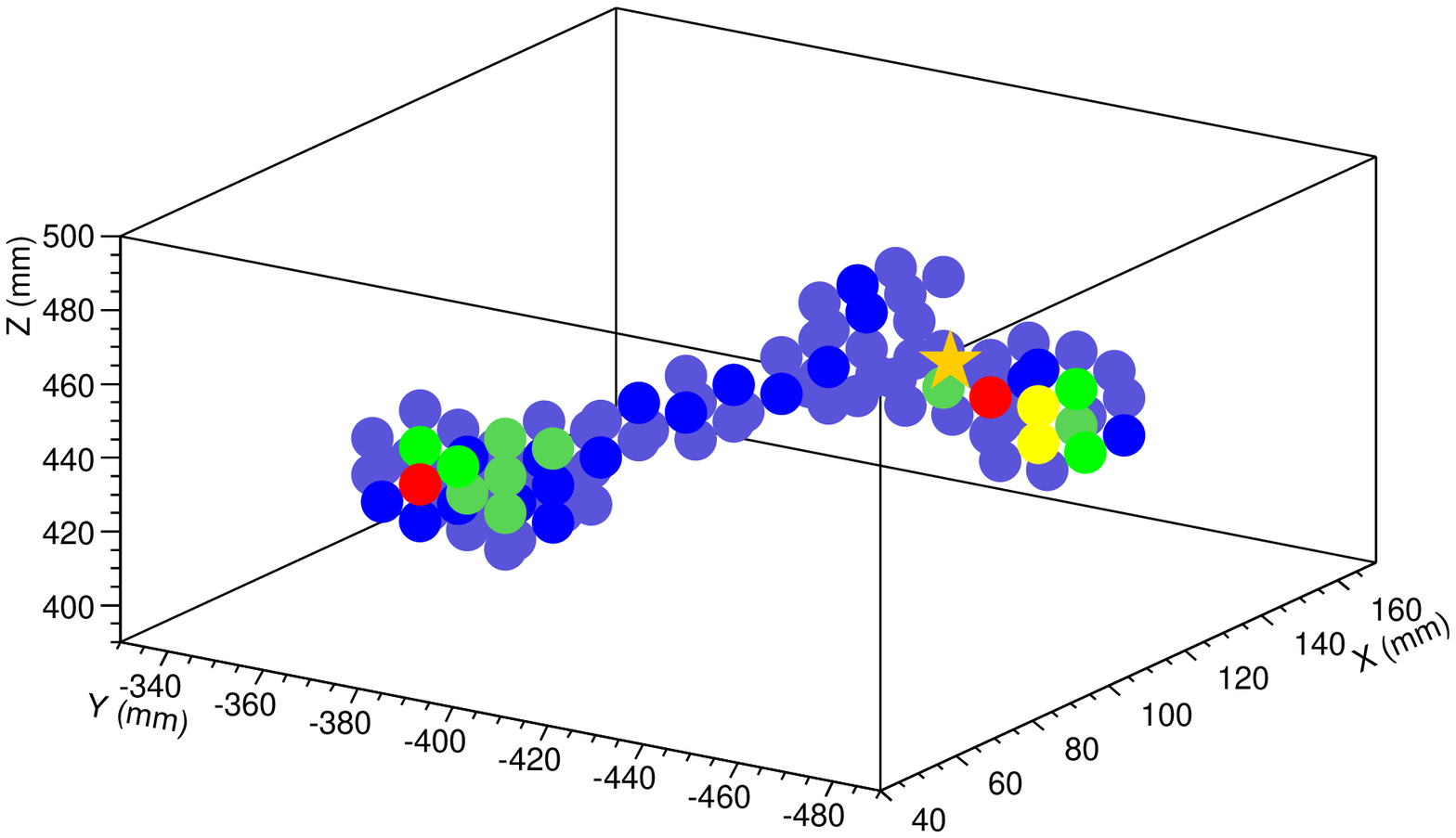}
                \label{figure6a}}
        \end{subfigure}
        \begin{subfigure}[]{
                \includegraphics[height=50mm,width=70mm]{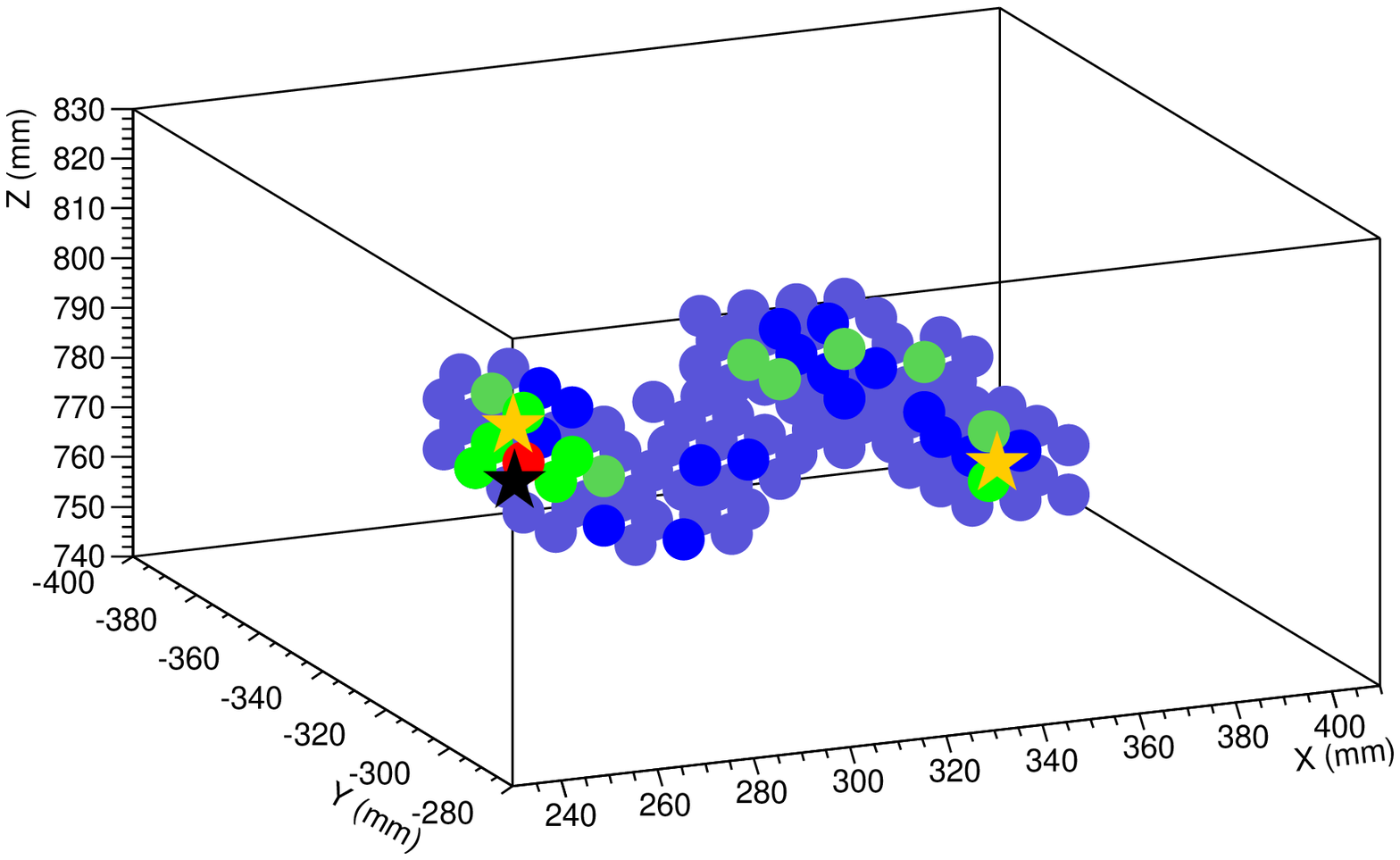}
                \label{figure6b}}
        \end{subfigure}
        \
        \begin{subfigure}[]{
                \includegraphics[height=60mm,width=75mm]{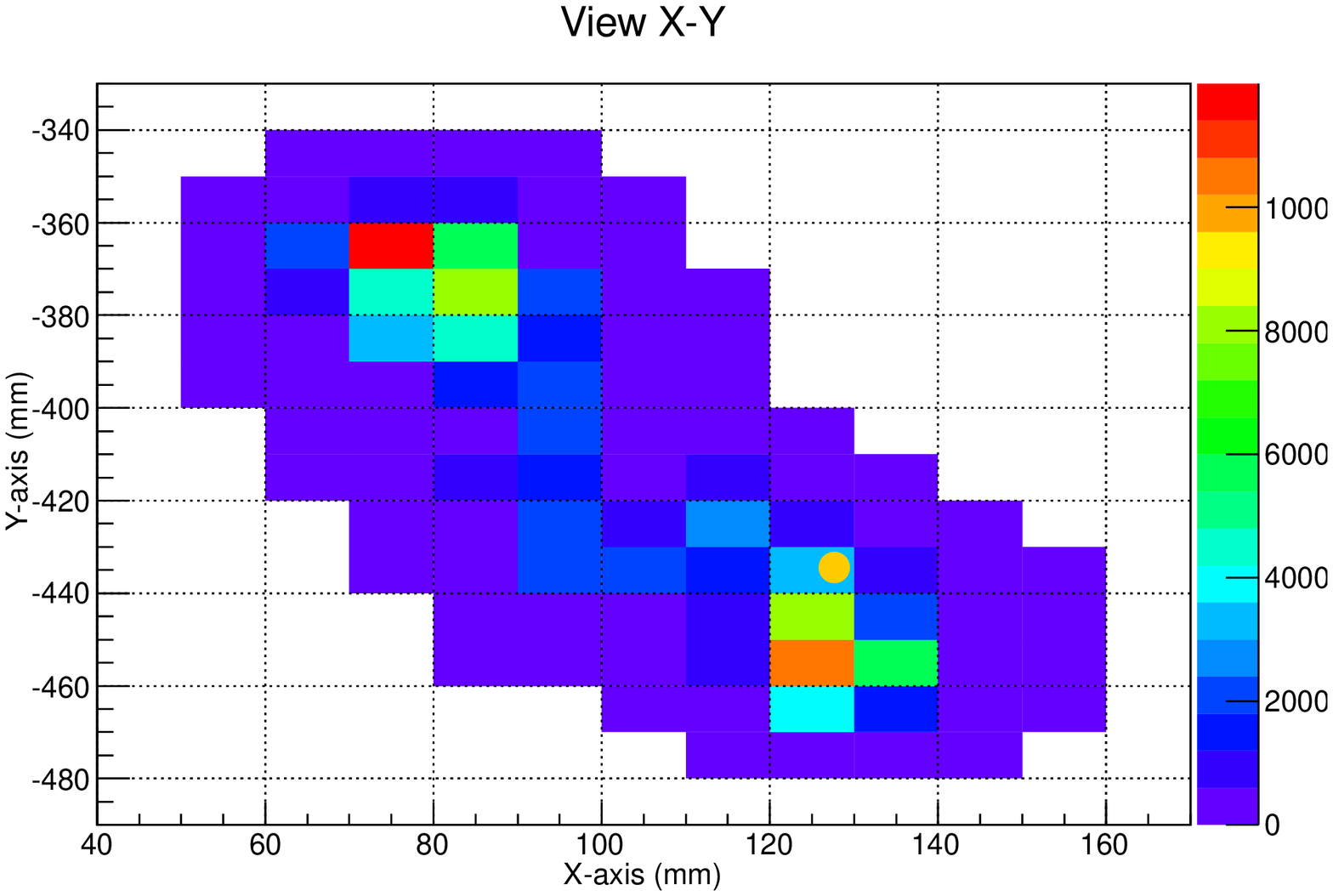}
                \label{figure6c}}
        \end{subfigure}
        \begin{subfigure}[]{
                \includegraphics[height=60mm,width=70mm]{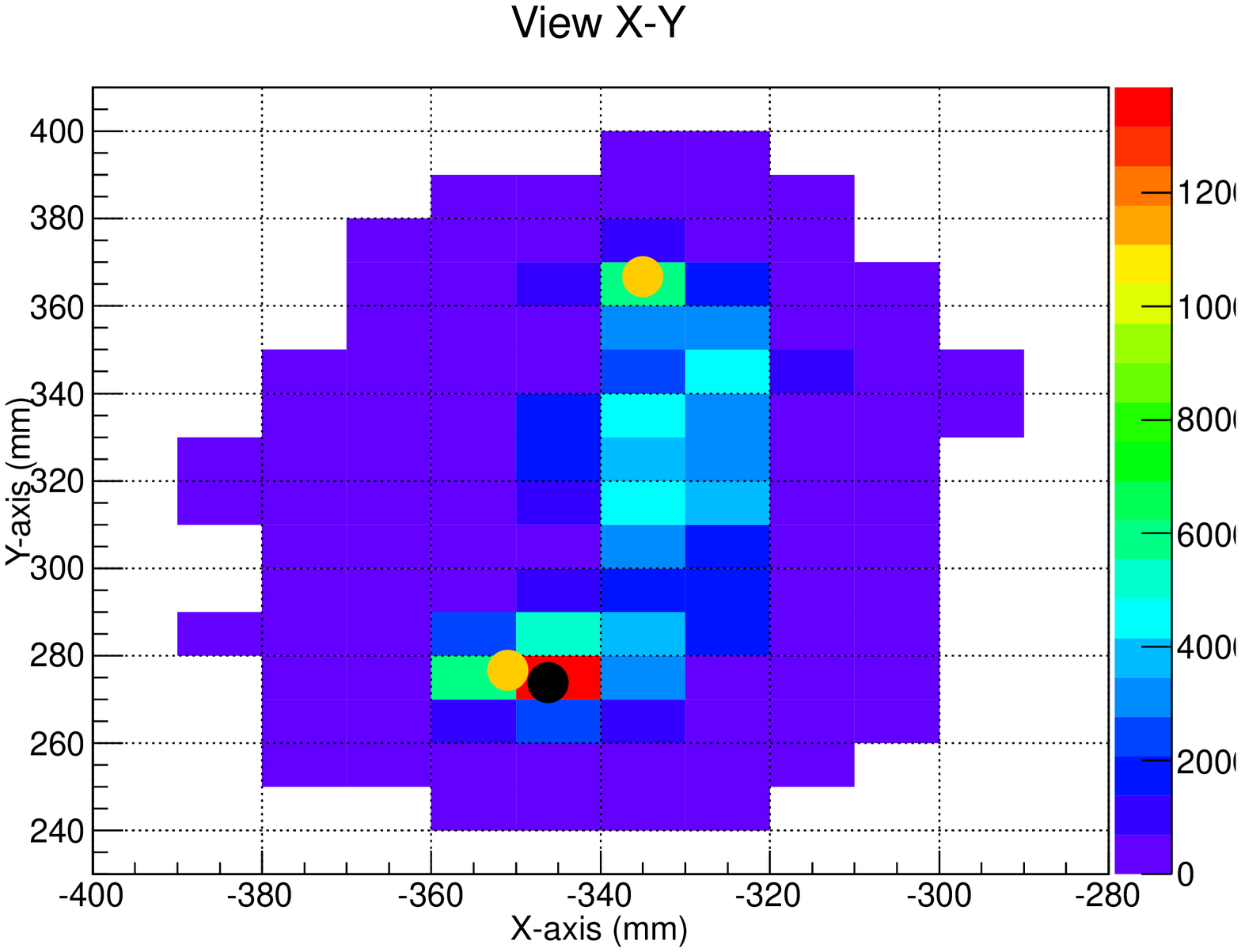}
                \label{figure6d}}
        \end{subfigure}
        \caption{ Three--dimensional representation (top) and two--dimensional projection (bottom) of a background ($^{214}$Bi ) event misidentified as signal event (left) and a $\beta\beta0\nu$ signal event misidentified as background event (right). On the left a $^{214}$Bi 2448\,keV photon has produced a secondary photon due to Compton interaction which has deposited its energy too close to the main track and originated a second blob. On the right, 3 low energy bremsstrahlung photons and 2 x-rays have interacted close to the track and originated extra energy depositions increasing the charge of one of the track's end.}\label{figure6}
\end{figure}

As already mentioned, most background events (coming from beta and gamma emissions from $^{208}$Tl and $^{214}$Bi contaminants) will produce multitrack topologies in the gas (see figure \ref{figure5}) and will be easily rejected. Beta decays from surface contaminations can  also be rejected by fiducialization. The most dangerous cases are when single tracks are produced in the fiducial volume. This can happen after photoelectric absorption of the $^{214}$Bi 2447.9\,keV photon emission, which falls right in the middle of the RoI (see figure \ref{figure6}). The 2614.5\,keV gamma emission of $^{208}$Tl can also produce single track of energy in the RoI if part of the event energy escapes the sensitive volume, either via Compton scattering or bremsstrahlung radiation.

The population of background events leaving a single track in the fiducial volume can be further constrained by looking for the characteristic two--blob feature of signal events. However, background events can still show higher energy depositions that mimic those from the Bragg peak in signal events, caused by secondary low energy radiation (x-rays or $\delta$-rays) or excessive straggling of the main track. These effects are also present in signal events and can affect their correct identification (and, therefore, the efficiency of the discrimination). Finally, the diffusion of the electronic cloud and readout digitization will affect in general the quality of the topological information, as only a somewhat blurry version of the physical track is available for analysis.

All the considerations here exposed constitute the rationale of the discrimination criteria. In the next section we describe quantitatively such criteria and the algorithm that implements them.

\section{The discrimination algorithms} \label{sec:Discrimination}

In the following we describe in detail the algorithm developed on the basis of the considerations discussed in the previous sections. It depends on a number of parameters, that will be introduced along the explanations of the algorithm below (they are all compiled in tables \ref{tab:track parameters}, \ref{tab:blob parameters1} and \ref{tab:blob para2}). The values for each of them have been chosen following diverse considerations (described in the following) to improve the discrimination of background and signal events, and so the performance of the algorithm. However, an overall optimization procedure of the whole set of parameters is still pending and remains for future work.

The algorithm is divided in three parts. The track method described in subsection \ref{subsec:Discrimination_tracks} identifies the number and charge of the tracks of the event and applies criteria on them. The blob method explained in subsection \ref{subsec:Discrimination_blobs} identifies the number and charge of the blobs of the main track and applies criteria on them. Finally, the fiducial method of subsection \ref{subsec:Discrimination_fiducial} applies a fiducial criterion.

\subsection{Discrimination based on number and energy of tracks}\label{subsec:Discrimination_tracks}

The algorithm identifies a track as a set of 3D–-pixels linked by a relation of proximity. This proximity relation can be defined as being adjacent or, in a more general way, as being closer than a given distance.  A distance of $d_p=10$\,mm around the pixel center has been chosen in this case to take into account adjacent pixels. The track identification is then based on a general mathematical graph algorithm described in \cite{Bollobas:1998bb, Gondran:1984mg}: each pixel of an event corresponds to a vertex of a mathematical graph and its relations to segments.  An initial point is then chosen and all vertices linked to it by one or successive segments are found. The process is repeated and all the connected points form a track.

Only pixels with a charge higher than a threshold $q_i>q_{\rm th}$ are considered due to two main reasons: a) in real life each pixel in the readout will have a threshold imposed by, e.g., a level of noise; and b) to reduce diffusion effects and have a better identification of separated clusters. To remain immune to the presence of pixels accidentally with no charge (due e.g.  to straggling or to technical problems in real data) a new charge $Q_i^{\rm t}$ is associated to each pixel in this method and defined as
 \begin{equation}\label{Qpixel}
    Q_i^{\rm t}=\sum_j q_j,
 \end{equation}
 where $j$ runs over all pixels closer than $r_{\rm t}=1$\,cm to pixel $i$ (i.e. adjacent pixels).  Only pixels with a $Q_i^{\rm t}\geq Q_{\rm th}^{\rm t}$ are finally taken into account, where we have taken $Q_{\rm th}^{\rm t}$ to be 45 electrons (around 1\,keV). Table \ref{tab:track parameters} summarizes the parameters defined for the method.

\begin{table}
\caption{\label{tab:track parameters} Parameters used in the track method and selected values. $q_{\rm th}$ is the threshold for the charge of a single pixel, $d_{\rm p}$ is the distance between two pixels to be considered as joined, $r_{\rm t}$ is the radius around a pixel to compute its charge in this method, $Q_{\rm th}^{\rm t}$ is the threshold in this charge for the pixel to be taken into account and $E_{\rm th}^{\rm t}$ is the threshold energy for \emph{short} tracks.}
\begin{indented}
\lineup
\item[]\begin{tabular}{@{}*{6}{l}}
\br
   Parameter          & $q_{\rm th}$  & $d_{\rm p}$& $r_{\rm t}$ & $Q_{\rm th}^{\rm t}$ & $E_{\rm th}^{\rm t}$\cr
\mr
  Value        & 45\,$e^-$  &10\,mm     &10\,mm 				 &45\,$e^-$ & 100\,keV\cr
\br
\end{tabular}
\end{indented}
\end{table}

\begin{figure}[tb]
\centering
\includegraphics[width=160mm]{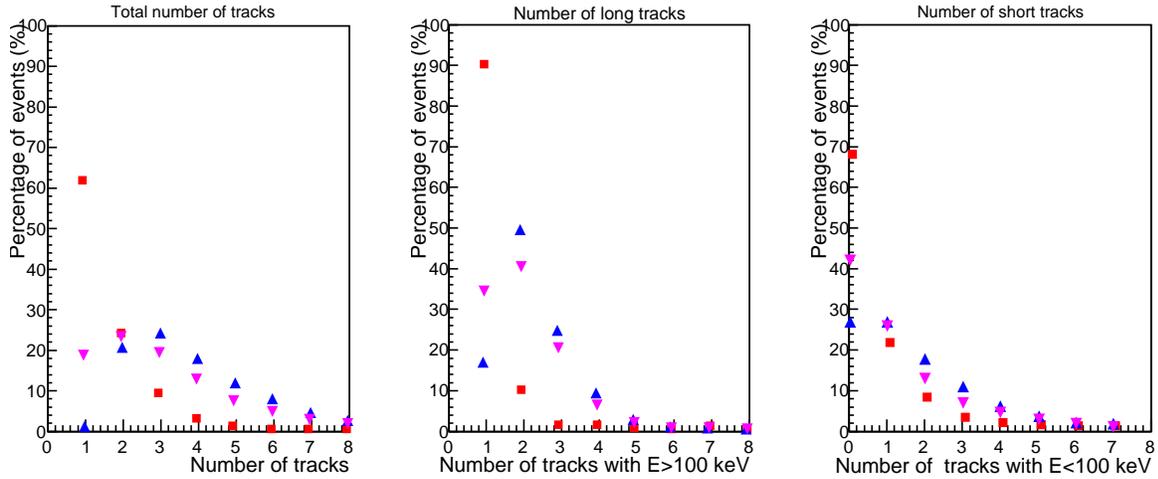}
\caption{Left: Number of identified tracks of any energy for events in the RoI generated by $\beta\beta0\nu$ (red squares), $^{208}$Tl (up blue triangles) and $^{214}$Bi (down purple triangle). As an example, the contaminations are chosen to emit from the field cage. Center: Number of \emph{long} (energies greater than 100\,keV) tracks. Right: Number of \emph{short} tracks (energies below 100\,keV).}
\label{figure7}
\end{figure}

As shown in figure \ref{figure7} (left), most of the background events have more than two tracks. However, only around a 60\% of signal events are single track, due to the emission of xenon x-rays (energies around 30\,keV) or bremsstrahlung photons (in more of the 80\% of the events) with typical energies below 100\,keV. These two effects produce a substantial reduction of signal efficiency if only single track events are selected. As observed in the plot at the centre of figure \ref{figure7}, if only the number of \emph{long} tracks (tracks with energies above $E_{\rm th}^{\rm t}=100$\,keV) were taken into account to reject an event, as done also in \cite{Wong:1993htw}, we could recover almost a 30\% of signal acceptance. However, the number of background events surviving this discrimination criterion would also increase. For this reason, the number of \emph{short} tracks has also been considered  (figure \ref{figure7}, right). After studying the effectiveness of the method for several possibilities, only those events with one \emph{long} track and up to one \emph{short} track will be selected as allowed signal topologies.

\subsection{Topological recognition of two blobs}\label{subsec:Discrimination_blobs}

\begin{figure}[bt]
        \centering
        \begin{subfigure}[]
                \centering
                \includegraphics[width=70mm]{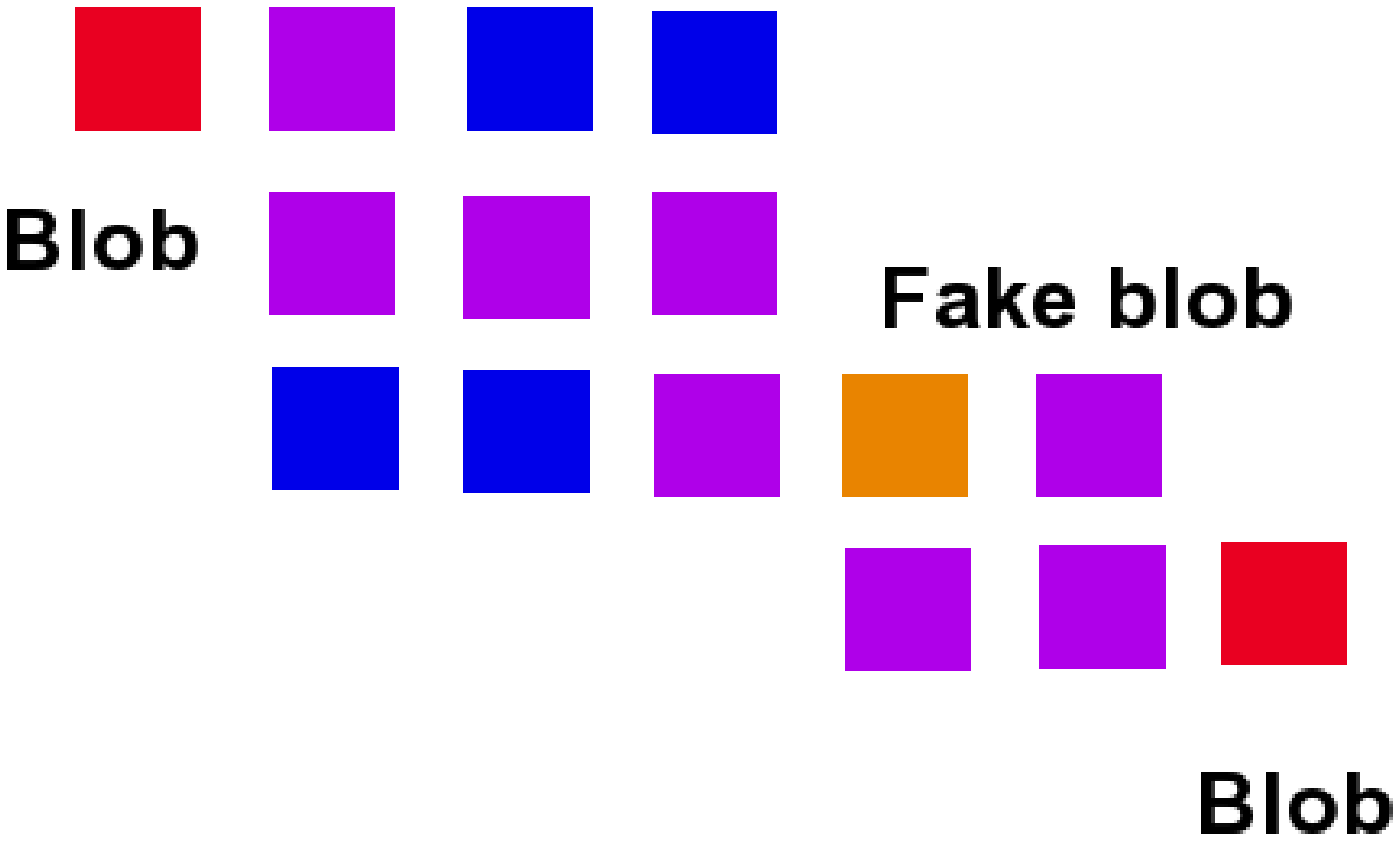}
                \label{figure8a}
        \end{subfigure}%
        ~ 
        \begin{subfigure}[]
                \centering
                \includegraphics[width=70mm]{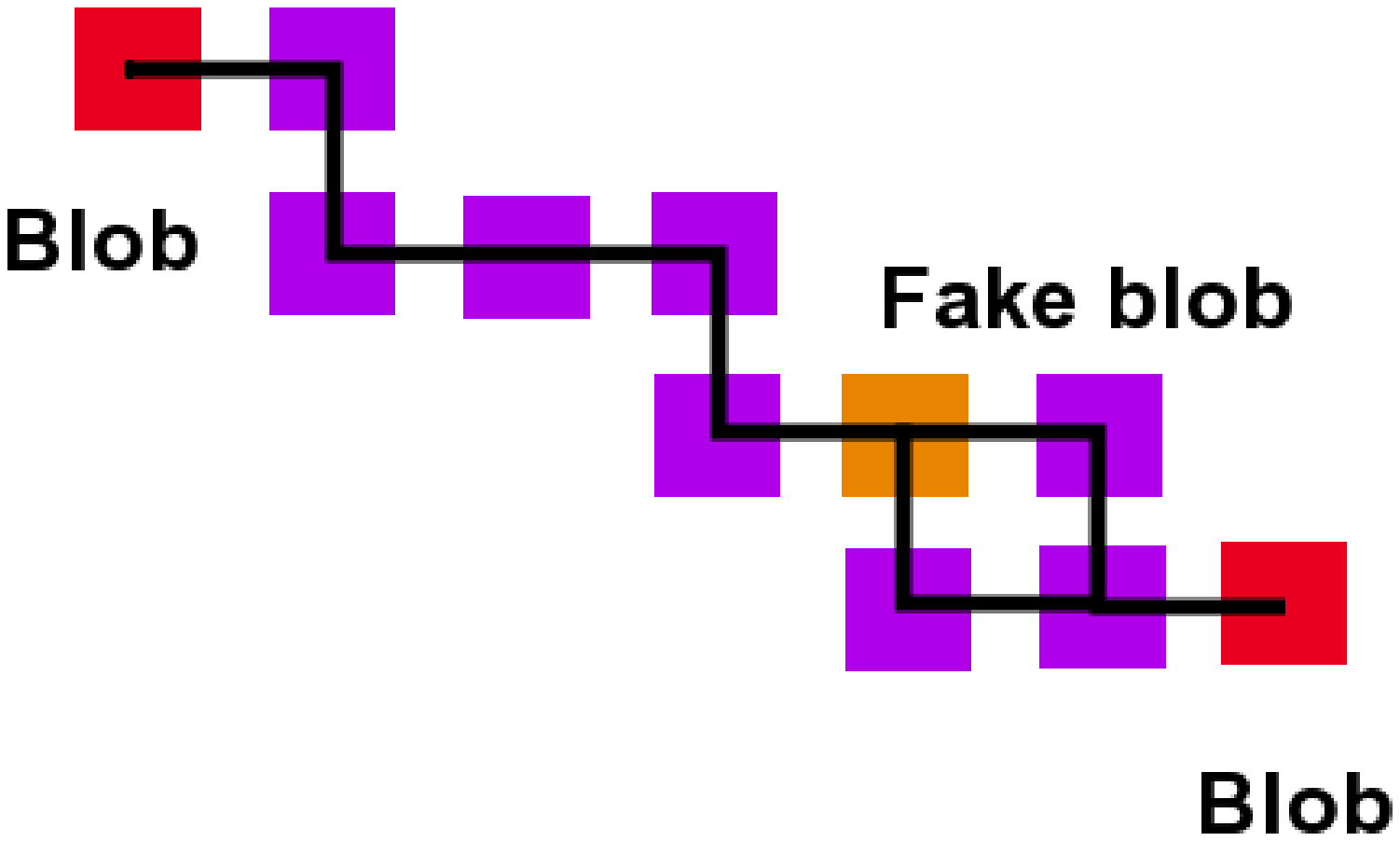}
                \label{figure8b}
        \end{subfigure}
        \
        \begin{subfigure}[]
                \centering
                \includegraphics[width=70mm]{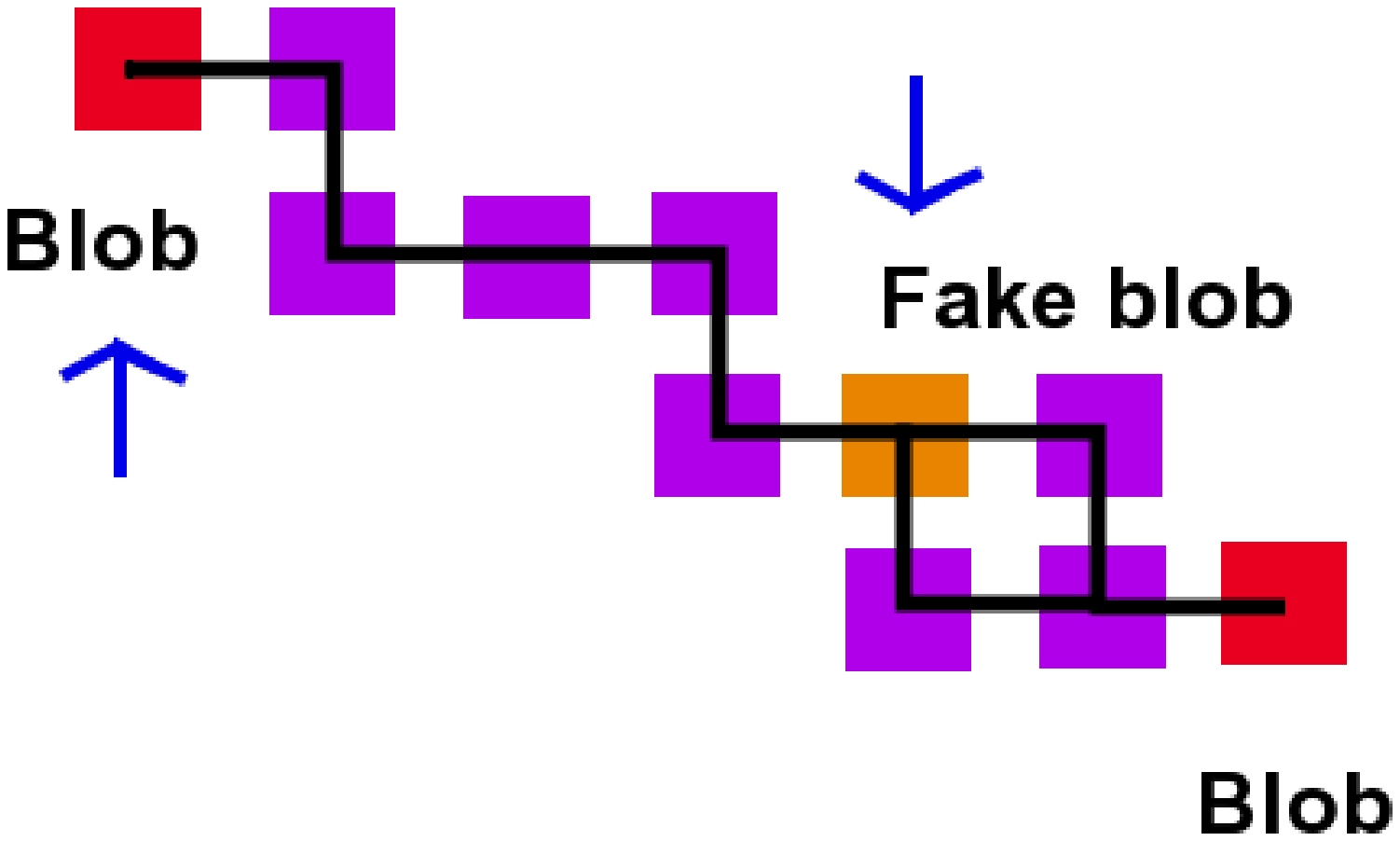}
                \label{figure8c}
        \end{subfigure}
        \begin{subfigure}[]
                \centering
                \includegraphics[width=70mm]{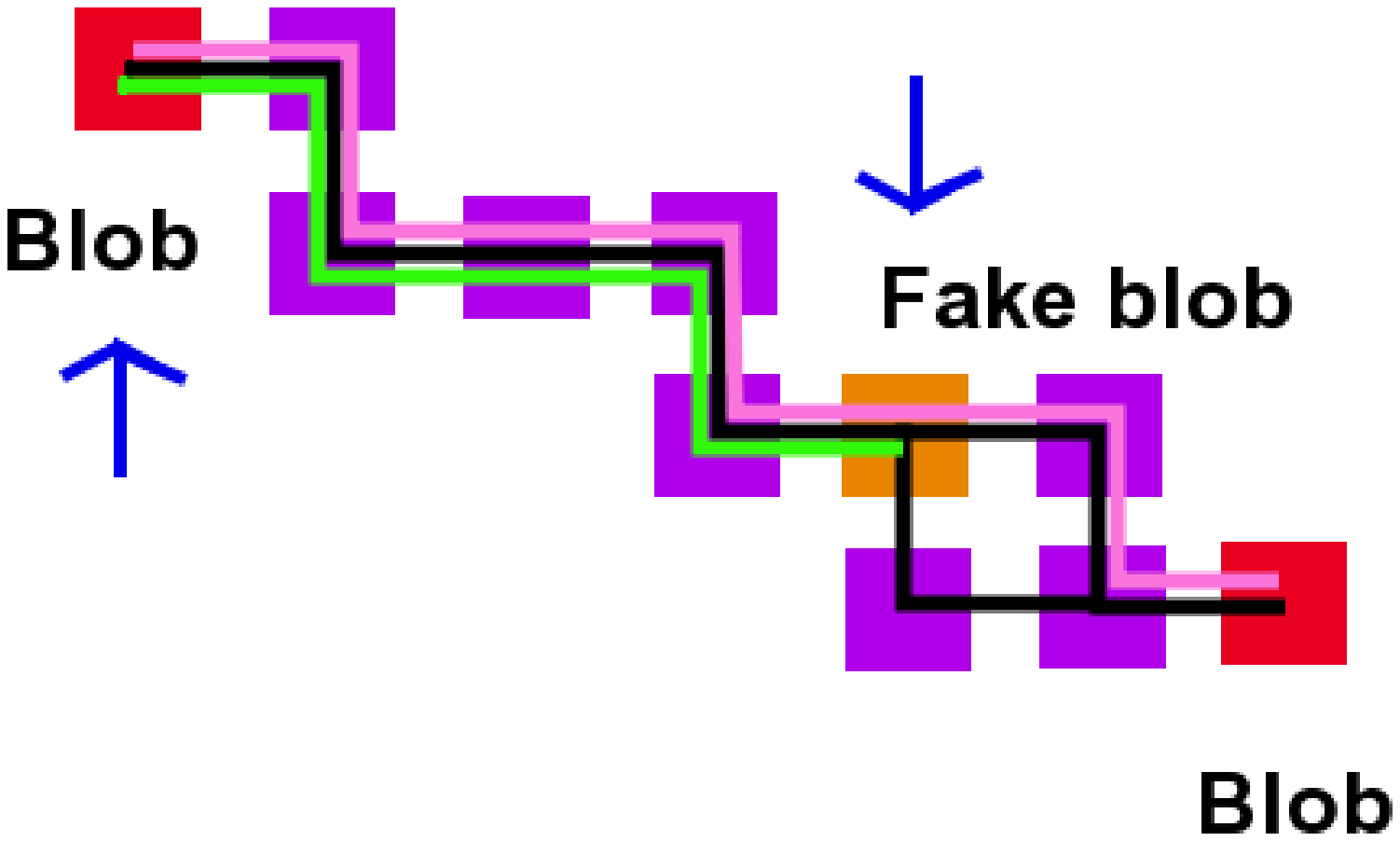}
                \label{figure8d}
        \end{subfigure}
        \caption{A dummy 2D event used to describe in the text the selection of the blob pixels. It consists of two real blobs (red squares), a fake one (orange square), normal pixels (magenta squares) and pixels with little charge (blue squares). From top left to bottom right: event pixels are identified with graph vertices and adjacent ones linked by segments; pixels with very low charge are removed to avoid twists; vertexes with the greatest charge (marked by arrows) are considered blob candidates;  two track--lines are drawn between the two blob candidates (pink line) or starting at one of the blob candidates and ending at some point with a charge accumulation (green line). As the first track--line is a 30\% longer than the second one, the first one is selected.}\label{figure8}
\end{figure}

This stage of the algorithm is only applied to the track of the highest energy in the event, identified in the previous stage. Its purpose is to locate the two blobs as well as the longest track--line (set of connected pixels of the track) that joins them both. Then several criteria are applied on the number and charge of the blobs found. This algorithm works in the following steps that are also graphically illustrated in figure \ref{figure8}.
\begin{enumerate}
\item For each pixel $i$ of the track a new charge $Q^{\rm b}_i$ is defined as the sum of the charges of all the pixels placed at a certain distance to it, $r_{\rm b}$, greater than the $r_{\rm t}$ distance used in the track method.
    \begin{equation}\label{PQequation}
        Q_i^{\rm b} =\sum_j q_j,
    \end{equation}
  where $j$ runs over all the pixels placed at a distance $d_{ij} < r_{\rm b}$. The distance $r_{\rm b}$ has been chosen to be 18\,mm.

 \item Blobs are defined as pixels with a charge higher than a certain threshold $Q_{\rm th}^{\rm b}=130$\,keV (almost 6000 electrons)
   \begin{equation}\label{bequation}
     Q_{i^\star}^{\rm b}> Q_{\rm th}^{\rm bc}.
   \end{equation}
 They are ordered according to their energy and the  first $N_{\rm C}=6$  of them labeled as \emph{blob candidates}.

\item  Any possible track--line (sequence of connected pixels within the track) between any two blob candidates is computed, as well as any track--line connecting each blob with any other pixel in the track. Only segments joining pixels with high charge depositions  ($Q_i^{\rm b}>Q_{\rm th}^{\rm b}$) are taken into account. The threshold $Q_{\rm th}^{\rm b}$ has been considered to be 150 electrons in the case of high diffusion and 75 electrons in low diffusion mixtures. This condition cleans part of diffusion effects.

\item  As the electron energy loss $dE/dx$ is higher at the end of its path, one--electron track should be longer than a two--electron track. Thus, the longest track--line between two \emph{blob candidate} and the longest track--line starting at a \emph{blob candidate} and ending at any other point are compared. If the second track--line is a $R_{\rm tl}=30\%$ longer than the first one, the second track--line is chosen, otherwise, the first one is considered. In such a way, track--lines where one of the ends is placed at the middle of the track are avoided. Finally, the two ends of the selected track--line are identified with the final \emph{two blobs} the algorithm was looking for. Notice that in some cases one of the ends may have not been labeled previously as \emph{blob candidate}.

\item  At this stage, a track--line and its two end--points have been identified. Charges $Q^1_{\rm blob}$ and $Q^2_{\rm blob}$ are  computed as the sum of the charges of all pixels inside of spheres of radius $r_{\rm blob}$ centered at both ends and assigned to the \emph{large blob} and to the \emph{small blob} respectively.  Only those events with similar charges at both ends and above a threshold energy ($Q_{\rm th}^{\rm blob}=440$\,keV) will be selected.

\begin{eqnarray}\label{BRatioequation}
  &Q^1_{\rm blob}>Q^2_{\rm blob}>Q_{\rm th}^{\rm blob}, \\ \nonumber \\
  &\frac{Q^{\rm blob}_2}{Q^{\rm blob}_1} \leq R_{\rm blobs},
\end{eqnarray}
with $R_{\rm{blobs}}$=2 in this work.

\begin{figure}[bt]
        \centering
        \begin{subfigure}[]
                \centering
                \includegraphics[width=75mm, height=55mm]{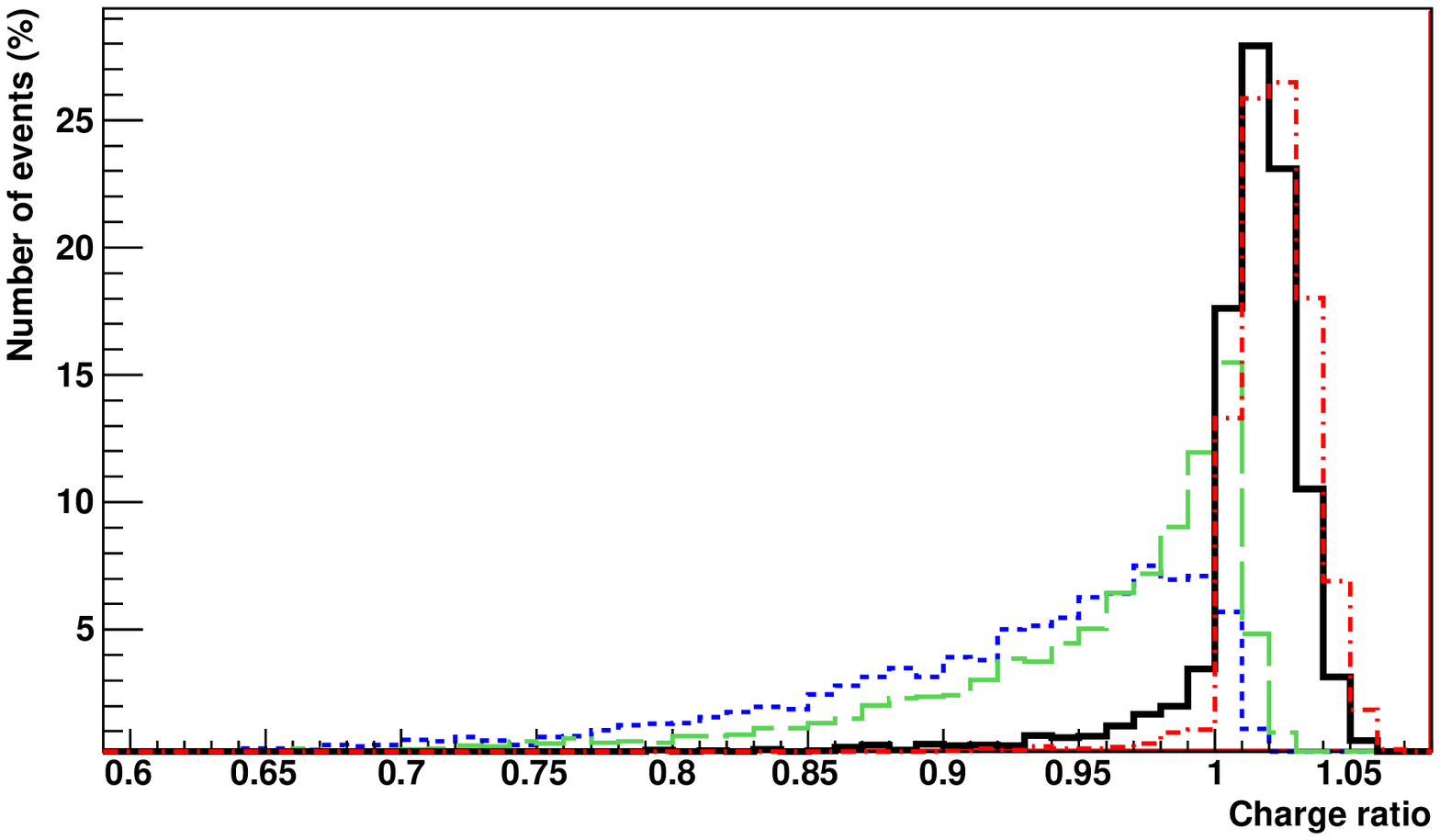}
                \label{figure9a}
        \end{subfigure}%
        ~ 
        \begin{subfigure}[]
                \centering
                \includegraphics[width=75mm, height=55mm]{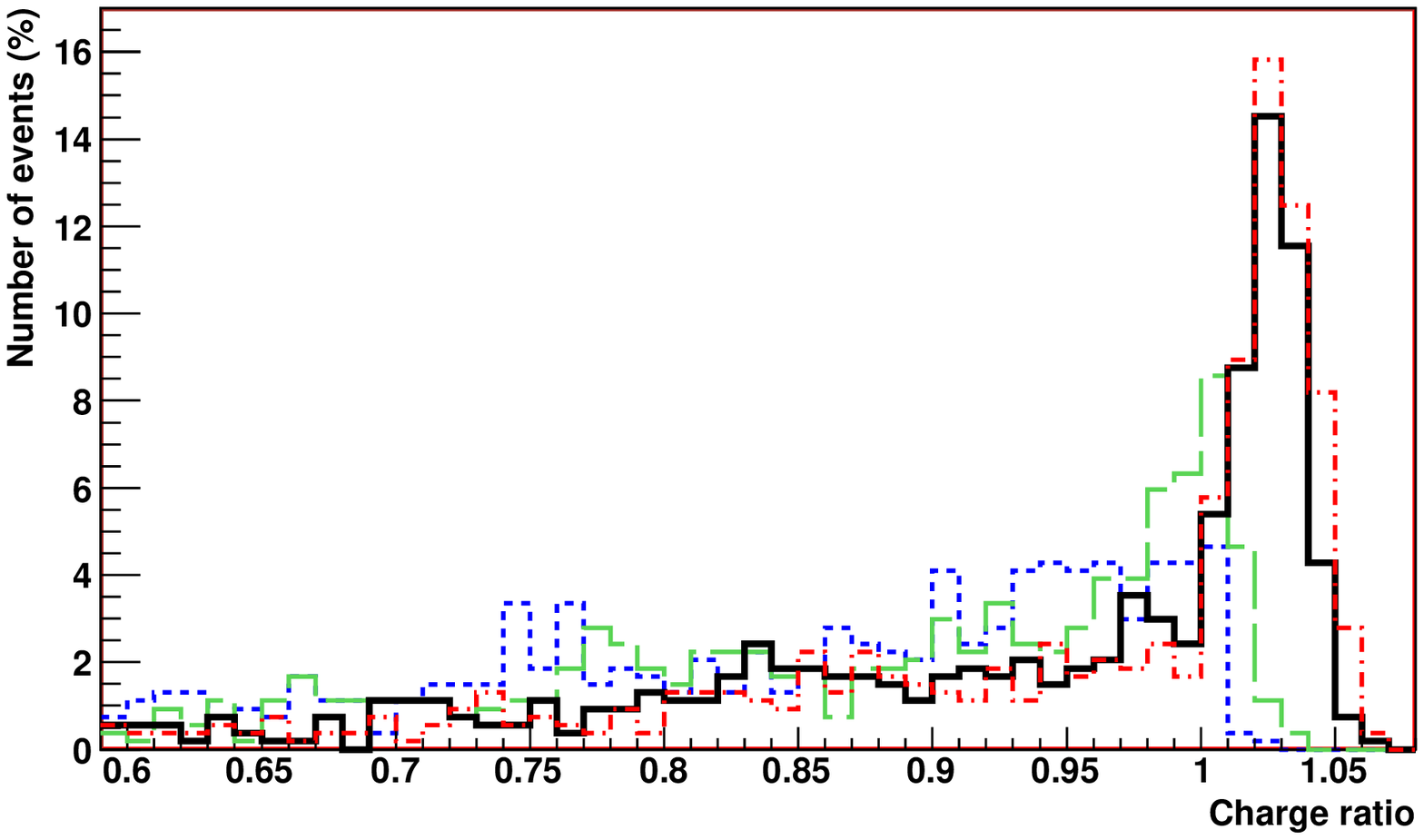}
                \label{figure9b}
        \end{subfigure}
        \
        \begin{subfigure}[]
                \centering
                \includegraphics[width=75mm,height=55mm]{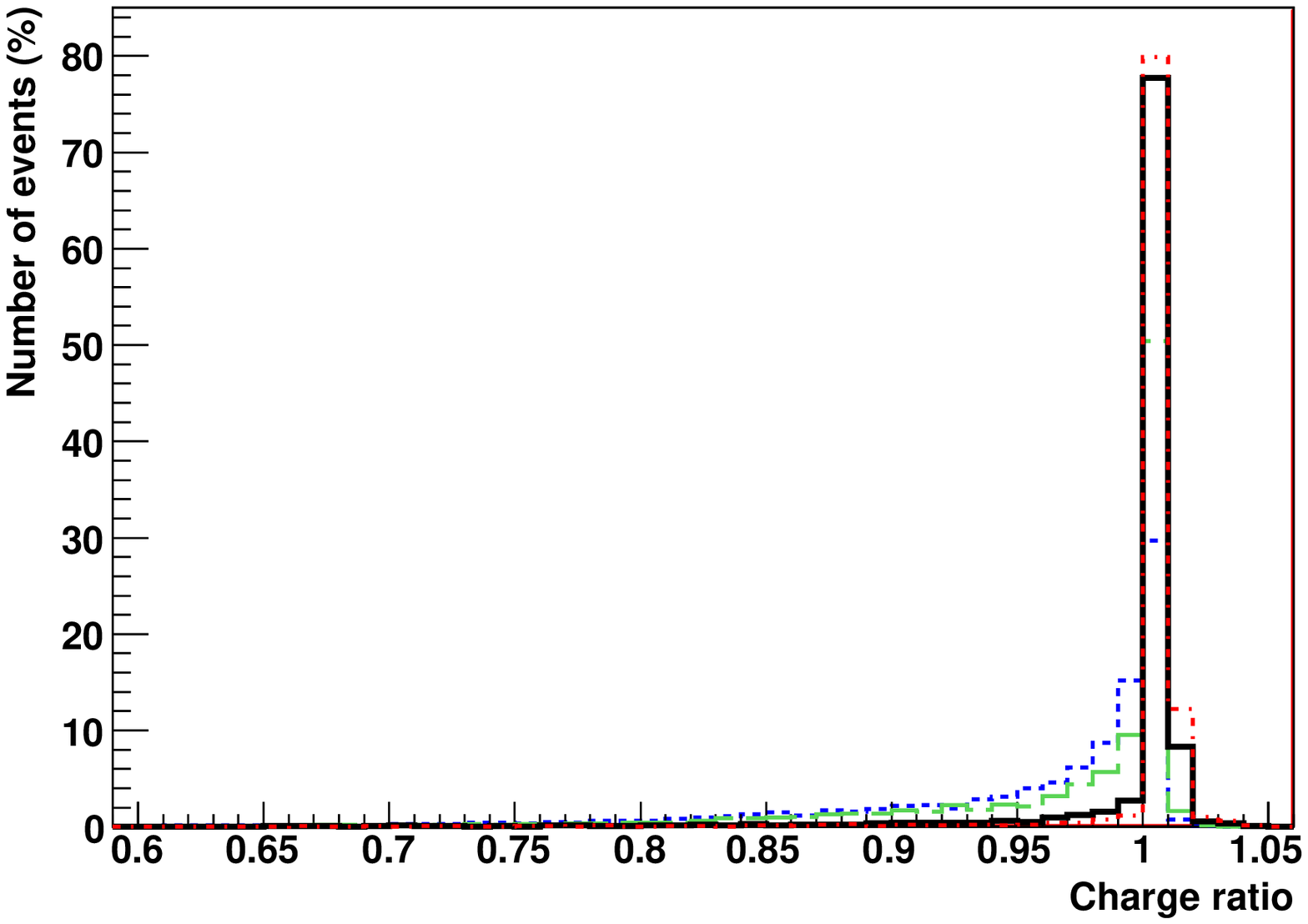}
                \label{figure9c}
        \end{subfigure}
        \begin{subfigure}[]
                \centering
                \includegraphics[width=75mm, height=55mm]{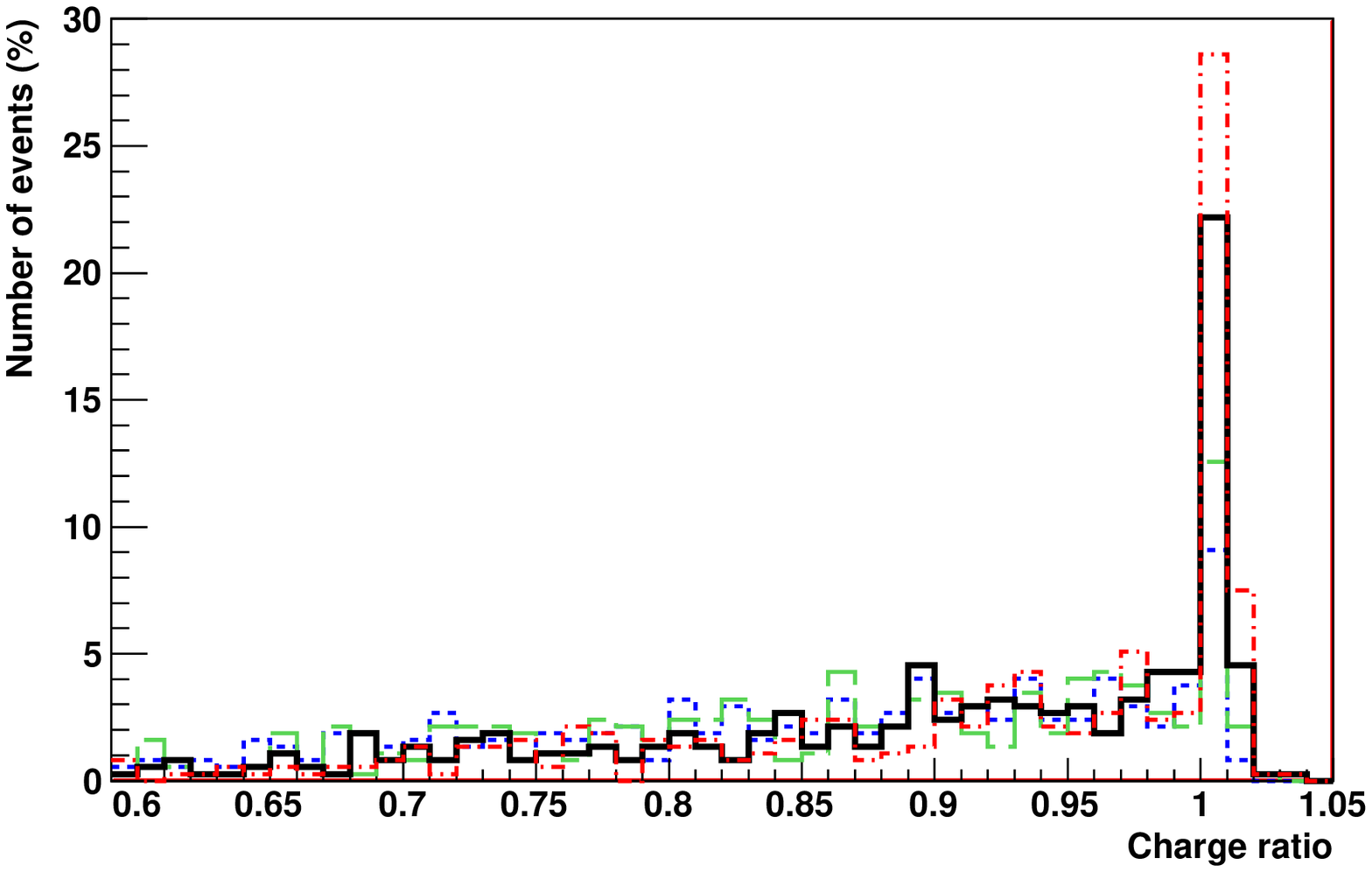}
                \label{figure9d}
        \end{subfigure}
        \caption{Figure shows the \emph{track coverage} in the case of xenon (plots on the left) and in the case of background events (plots on the right). Upper plots correspond to pure xenon gas detector and lower ones to the use of a low diffusion mixture. On each plot the ratios for different values of the radius $r_{\rm lt}$ of the \emph{lego tube} are shown: solid line for 30\,mm (chosen value), dashed line for 15\,mm, long dashed line for 18\,mm and dotted line for 36\,mm. As can be seen, most of the signal events have a charge ratio around 1, accentuated in the case of a low diffusion mixture. Values above 1 mean that extra charges from pixels below the threshold $q_{\rm th}$ used to separate tracks or from smaller tracks have been enclosed inside the \emph{lego tube}.  }\label{figure9}
\end{figure}

\item For a small fraction of events, the final two blobs identified by the algorithm do not correspond to the real ones. In those cases, the track--line found out by the algorithm follows only part of the real topology, leaving the other part outside. To veto these cases, an additional variable labeled \emph{track coverage} has been defined as the ratio between the charge contained in a given \emph{lego}--tube around the calculated track--line (a $r_{\rm lt}=30$\,mm radial distance for the tube has been chosen, see figure \ref{figure9}) and the total charge of the track. Then, it is required that more of the $R_{\rm cv}$ of the charge be inside this tube ($R_{\rm cv}=95\%$ for pure xenon and $R_{\rm cv}=98\%$ in the case of low diffusion Xe gas mixture). It can be seen how the use of a low diffusion mixture improves the pattern recognition.

\end{enumerate}

The first parameters of the method ($r_{\rm b}$, $Q_{\rm th}^{\rm b}$, $Q_{\rm th}^{\rm bc}$, $N_{\rm C}$ and $R_{\rm tl}$, summarized in table \ref{tab:blob parameters1}) have been chosen to optimize the effectiveness of the method to find the real blobs of the event. For most cases, the distance between the real and the reproduced blob position is less than 2 times the pixel length.

\begin{table}
\caption{\label{tab:blob parameters1} Parameters used in the blob method and their selected values to find the longest track--line. $r_{\rm b}$ is the radius around a pixel to compute its charge in this method, $Q_{\rm th}^{\rm b}$ is the threshold in charge for a pixel to be joined to other by segments, $Q_{\rm th}^{\rm bc}$, the threshold in the charge to be considered as a \emph{blob candidate}, $N_{\rm C}$ is the number of blobs candidates, $R_{\rm tl}$, the maximum ratio between the longest track--line with two blobs and just one to choose the second one. In brackets values for low diffusion mixtures in case they are different from those of pure xenon.}
\begin{indented}
\lineup
\item[]\begin{tabular}{@{}*{6}{l}}
\br
   Parameter    &$r_{\rm b}$  &$Q_{\rm th}^{\rm b}$ & $Q_{\rm th}^{\rm bc}$ & $N_{\rm C}$ & $R_{\rm tl}$\cr
\mr
  Value         & 18\,mm  &150\,$e^-$	&	 130\,keV	 &6 & 30\%  \cr
                 &   &(75\,$e^-$) 		&	 	 & & \cr
\br
\end{tabular}
\end{indented}
\end{table}

 Table \ref{tab:blob para2} summarizes the second part of parameters used in the blob method ($r_{\rm blobs}$, $Q_{\rm th}^{\rm blobs}$, $R_{\rm blobs}$, $r_{\rm lt}$, $R_{\rm cv}$). To select the parameter values, it has been looked for ranges of these parameters which maximizes the figure of merit $\sim \epsilon\sqrt{F}$, where $\epsilon$ is the signal efficiency and $F$ is the rejection factor due to a specific discrimination criterion. Among values given equal figure of merit, we have favored the highest signal acceptance. As an example, once the radius $r_{\rm lt}$ has been fixed to 30\,mm, ranges of $Q_{\rm th}^{\rm blob}$ from  440\,keV to 572\,keV  and $R_{\rm cv}$ from 95\% to 98\%  are maximize the figure of merit in the case of pure xenon, but 440\,keV for $Q_{\rm th}^{\rm blob}$ and 95\% for $R_{\rm cv}$ present the highest signal acceptance. As already mentioned, an optimization of the whole set is foreseen in future work.

\begin{table}
\caption{\label{tab:blob para2} Parameters used in the blob method and their selected values to discriminate between one--blob and two--blobs track--lines. $Q_{\rm th}^{\rm blob}$ is the threshold energy for any of the two end blobs of selected track--line which have been calculated in a radius $r_{\rm blob}$, $R_{\rm blobs}$, the maximum ratio between both end blobs, $r_{\rm lt}$, the radius around the track--line to estimate the charge inside, and, finally $R_{\rm cv}$ is the minimum charge ratio which has to be included inside this radius versus the total charge. In brackets values for low diffusion mixtures in case they are different from those of pure xenon.}
\begin{indented}
\lineup
\item[]
\begin{tabular}{@{}*{6}{l}}
\br
   Parameter    & $r_{\rm blobs}$ &$Q_{\rm th}^{\rm blobs}$ & $R_{\rm blobs}$ & $r_{\rm lt}$ & $R_{\rm cv}$\cr
\mr
  Value         & 30\,mm &440\,keV &  2& 30\,mm & \095\% \cr
                  && & & & (98\%)\cr
\br
\end{tabular}
\end{indented}
\end{table}

\subsection{Fiducial criterion} \label{subsec:Discrimination_fiducial}

Among all the background sources, we need to deal with gammas, but also with beta particles produced on surfaces close to the gas sensitive volume. To exclude these events the outermost part of the active volume ($r_{\rm veto}=1$\,cm width) is treated as an active veto, i.e., events depositing energy in this region are rejected. Experimentally it is easy to identify events close to the field cage walls: pixels in the border of the tracking plane will contain charge in this case. Events produced close to the cathode or the anode
can be identified using the $t_0$ of the event. The same situation occurs for those events near the anode. This criterion has been applied after the previous ones to study its real impact on background rejection and signal efficiency since part of these \emph{surface} events will have already been rejected, mainly after the one track method.

\section{Background rejection factor and signal efficiency after discrimination criteria} \label{sec:results}

In this section the results of the application of the discrimination algorithms described in section \ref{sec:Discrimination} on background and signal simulated events will be presented. Background rejection factors have been computed separately for both $^{208}$Tl and $^{214}$Bi contaminations, as well as for each of the four representative volumes defined in section \ref{sec:sim}: cathode, readout plane, field cage and lateral part of the vessel.

Our results are summarized in several tables in this section for both, background rejection factor and signal efficiency, in the case of pure xenon, high diffusion (HD), and a low diffusion (LD) Xe mixture. Two sets of tables are presented for each of the cases: a) the net reduction in both gases, to be commented below; and b) the relative effect of each selection criteria applied subsequently  to be detailed in subsections \ref{subsec:results_xenon} and \ref{subsec:results_xenon_TMA};

Focusing our discussion on efficiencies (table \ref{tab:efficiency}), it can be appreciated that only around a 70\% of the $\beta\beta0\nu$ events are fully contained in the detector and deposit their energy in the chosen RoI. That is just detector geometry dependent and, therefore, the same for both gases. In the Gothard experiment, a much smaller TPC, this value was around 30\% \cite{Wong:1993htw} for a RoI between 2 and 3\,MeV.

The last two columns of the table \ref{tab:efficiency} give information about final efficiencies. They show similar numbers for both gases: over the 40\%. That makes almost a 60\% of efficiency if just RoI events are taken into account, a bit below the value of 74\% given by Gothard experiment, but where the inefficiency was estimated as only due to events where one of the electrons had not enough energy to produce a blob. However, considering also the geometrical factor, Gothard had estimated a 22\% of overall efficiency to be compared to the 40\% estimated in the present work.

\begin{table}
\caption{\label{tab:efficiency} Number of signal events (in \%) in the RoI around the $Q_{\beta\beta}$ (first column) and final efficiency after all the discrimination algorithms have been applied in the case of  pure xenon (second column) and a low diffusion xenon mixture (last column).}
\begin{indented}
\lineup
\item[]\begin{tabular}{@{}*{4}{l}}
\br
        & RoI events        & Final eff.(HD) &  Final eff.(LD) \cr
\mr
$^{136}$Xe$_{\beta\beta0\nu}$              &71.1 $\pm$ 0.2     &40.7 $\pm$ 0.2 				 &40.0 $\pm$ 0.2\cr
\br
\end{tabular}
\end{indented}
\end{table}

 In table \ref{tab:AccfacTotXe}, background reduction factors (i.e. number of events surviving the cuts per simulated one) are shown. Values for events in the RoI and for events after all the cuts are given.

\begin{table}
\caption{\label{tab:AccfacTotXe} Background reduction factors for the different background sources in a RoI around the $Q_{\beta\beta}$ (first column), and after the application of all selection criteria in the case of pure xenon (second column) and in the case of a low diffusion xenon mixture (last column).}
\begin{indented}
\lineup
\item[]\begin{tabular}{@{}*{5}{l}}
\br
	Origin & Isotope &   RoI events &Final ev. (HD) & Final ev. (LD)\\
  \ns
   & & ($\times 10^{-3}$) & ($\times 10^{-6}$)  & ($\times 10^{-6}$) \\
\mr
\lineup

Lateral & $^{208}$Tl              &0.4  							&\07.0$\pm$0.6              &2.5$\pm$0.4 \\
 vessel& $^{214}$Bi			 &0.03            							  &\01.5$\pm$0.1               &0.5$\pm$0.1\\[1mm]
  											
Field & $^{208}$Tl    &2.0    							  &13.4$\pm$0.1               &7.2$\pm$0.9 \\
  cage		 & $^{214}$Bi			 &0.2   		  &\04.1$\pm$0.3               &1.7$\pm$0.1\\[1mm]

 \multirow{2}{*}{Readout}& $^{208}$Tl    &2.7    								&\05.4$\pm$0.7               &2.1$\pm$0.5\\
  		& $^{214}$Bi	 &1.4    &\00.9$\pm$0.2               &0.2$\pm$0.1\\[1mm]

 \multirow{2}{*}{Cathode} & $^{208}$Tl      &2.2   							  &\07.4 $\pm$ 0.8               &4.4$\pm$0.6 \\
  	& $^{214}$Bi							   &1.0   								 &\01.1$\pm$0.2               &0.4$\pm$0.1\\

\br
\end{tabular}
\end{indented}
\end{table}

 The first column of table \ref{tab:AccfacTotXe} reveals that only a small fraction of the simulated events deposit their energy in the chosen RoI: $10^{-3}$ for events originated close to the gas and smaller fractions for events coming from farther volumes due to attenuation and smaller solid angles; then, the farther volume, the lower fraction will pass. The reason why the RoI event fraction is larger for the $^{214}$Bi contamination on surfaces facing the sensitive volume is due to the fact that additional beta electrons can reach the active volume and deposit part of their energy the RoI.

The second and third columns of table \ref{tab:AccfacTotXe} exhibit an additional reduction of at least three orders of magnitude due to the use of topological discrimination, a factor 3 better in the case of a mixture with a lower diffusion since pattern recognition improves (discussed in subsection \ref{subsec:results_xenon_TMA}). This reduction is slightly worse for events coming from the cathode because the distance to the readout plane plays also an important role in diffusion. For events in the RoI, the background reduction factors depend very much on the origin of the contamination being much more effective for surface contaminants (cathode and readout cases) due mainly to the presence of electrons (more relevant in the case of the $^{214}$Bi), than for a volume contamination (case of the vessel). In this last case, however, it is easier to reject $^{208}$Tl events than those caused by $^{214}$Bi due mainly due mainly to the multi--track topology of $^{208}$Tl events, as discussed later.

\subsection{High Purity Xenon Gas: a high diffusion medium} \label{subsec:results_xenon}

\begin{table}
\caption{Surviving events(in \%) of the different sources of background simulated after the successive application of the selection criteria in the RoI for a 1 cm--length pixelized detector in a xenon at 10 bar. Each column shows the effect of a given cut on the event population selected by the previous one.
\label{tab:rejfaCutsXe}}
\begin{indented}
\lineup
\item[]\begin{tabular}{@{}*{7}{l}}
\br 		                              &            &              &                 &\centre{3}{Fiducial method} \\
\ns
&&&& \crule{3} \\
 Origin & Isotope  & Track method        & Blob method            &Lateral    &Bottom        &Top\\
\mr
\lineup
Lateral  & $^{208}$Tl   &  \07.0$\pm$0.3     &41.9$\pm$3.8         &68.9$\pm$8.1          &100.0$\pm$12.8   &\098.4$\pm$12.6\\
vessel  & $^{214}$Bi	 &  14.1$\pm$0.5    &43.1$\pm$2.7         &82.1$\pm$6.4          &\099.7\0$\pm$8.2   &\099.0\0$\pm$8.1\\[1mm]
Field & $^{208}$Tl   &  \04.8$\pm$0.1     &43.5$\pm$2.0   &33.9$\pm$2.6   &\097.8\0$\pm$9.3  &\097.7\0$\pm$9.4\\
 cage & $^{214}$Bi	&  41.9$\pm$0.7    &39.4$\pm$1.0         &12.7$\pm$0.8          &\099.6$\0\pm$8.8     &\098.1\0$\pm$8.7\\[1mm] 	 
 \multirow{2}{*}{Readout}& $^{208}$Tl   &  \03.2$\pm$0.1 &40.7$\pm$2.6 &90.8$\pm$7.1   &\016.8\0$\pm$2.5     &100.0$\pm$19.4\\
  	& $^{214}$Bi&  51.6$\pm$0.5    &36.6$\pm$0.6         &88.6$\pm$1.7          &\0\00.4\0$\pm$0.0     &100.0$\pm$32.4\\[1mm]
 \multirow{2}{*}{Cathode} & $^{208}$Tl  &  \03.8$\pm$0.1     &56.9$\pm$3.2     &80.0$\pm$5.3  &100.0\0$\pm$7.0     &\019.1\0$\pm$2.4\\
                        & $^{214}$Bi &65.7$\pm$1.1    &  34.9$\pm$0.4      &76.6$\pm$1.5    &100.0\0$\pm$2.0     &\0\00.6\0$\pm$0.1\\
\br
\end{tabular}
\end{indented}
\end{table}

Table \ref{tab:rejfaCutsXe} summarizes the relative background reduction factors for each of the background sources. Only events inside the RoI will are taken into account here. As already mentioned, each discrimination method is applied to events which have passed previous cuts. After the first selection, based on track method, the number of events surviving is around 3--7\% for the $^{208}$Tl and  from $14\%$ up to $66\%$ for the $^{214}$Bi, depending on their origin. Three points to are to remark here: a) the systematically better rejection factor of the track method observed for  $^{208}$Tl events is due to the fact that they produce multi--track topologies (due to Compton or bremstrahlung) more often than $^{214}$Bi events; b) in the case of surface contaminations (cathode, readout) low energy electrons, emitted in coincidence with photons, reach the active volume and produce multi--track events, but, on the contrary, electrons from high energy $\beta$ decays of $^{214}$Bi increases the number of single--track events;and c)in the case of $^{208}$Tl, the track method is slightly less effective for events originated far from the gas since they may have suffered Compton interactions in materials resulting in one--track events in the detector.

 The effect of the topological recognition is quite similar for all contributions of background independently of origin and isotope, as expected since it is applied on the main track. The surviving events after the application of this method with respect to the previous one is of the order of $40\%$. The slight differences may be attributed to two reasons: a) events due to $\beta$ emissions ($^{214}$Bi on surfaces) have less secondary tracks, so they have passed the track criterion,  but most of them show clearly one blob at the end far from the surface and no extra charge at the end close to the surface, so they tend to be better rejected by the blob method; and b) diffusion joining tracks affects more to events produced at longer drift distances and  makes more difficult the topological recognition for events from the cathode. That is commented later in next subsection (\ref{subsec:results_xenon}) when reporting on a low diffusion Xe mixture data.

The fiducial rejection criteria has a different effect depending on the origin of the background, playing a more important role for surface contamination, as pointed out in subsection \ref{subsec:Discrimination_fiducial}, while for volumes far from the gas, its reduction is purely statistical. It has been applied separately in the three directions: first in the laterals and then in the bottom and top. As shown in table \ref{tab:rejfaCutsXe} the rejection of events firing the outermost readout pixels (lateral cut) has the most important impact, of the order of a $20\%$, since it is the largest surface. It has to be noted, however, that in the case of volumes close to the gas (field cage) or surface contaminations (cathode and readout planes) the effect of the corresponding discrimination criterion is higher since beta emission can reach the gas and be rejected. They accompany the gamma  emission in the case of $^{208}$Tl and are part of the high energy beta spectrum for $^{214}$Bi.  Here the rejection factor is higher than $99\%$. It is interesting to note that apart from these surface beta emissions the effect of the fiducial cut is otherwise rather modest. This result is particularly interesting in order to consider operation without $t_0$ determination, something useful in the context of certain gas mixtures with quenched scintillation.

A more detailed analysis of surviving events shows that most of them have been identified as signal due to the presence of high energy ($>$100\,keV) secondary  photons interacting too close to the main track and producing a second blob. Roughly, half of them are due to a Compton interaction, another fraction (around a 10\%) to high bremsstrahlung emissions, and the remaining to low radiation and x-rays photons, straggling, or failures of the recognition algorithms.

\begin{table}
\caption{Relative efficiency of the simulated $\beta\beta0\nu$ events after the successive application of selection criteria on events depositing energy in the RoI. Each column shows the percentage of surviving events after a cut on events which have passed the previous one. Considered a 1 cm---length pixelized detector in pure xenon gas at 10\,bar.
\label{tab:efficiencyCutsXe}}
\begin{indented}
\lineup
\item[]\begin{tabular}{@{}*{7}{l}}
\br
  		                              &            &              &                 &\centre{3}{Eff. Fiducial vetoes} \\
\ns
&&&& \crule{3} \\
 Origin & Isotope  & Eff. Track        & Eff. Topology            &Lateral    &Bottom        &Top\\
\mr
\lineup
      Target &$^{136}$Xe                   &77.5$\pm$0.4   &85.4$\pm$0.5       &88.5$\pm$ 0.5				   &99.1 $\pm$0.6    &98.6$\pm$0.6    \\
\br
\end{tabular}
\end{indented}
\end{table}

Regarding signal events, table \ref{tab:efficiencyCutsXe} shows the impact of cuts on $\beta\beta0\nu$ signal once events in the RoI have been selected. The most important reduction in efficiency is due to the selection of events with just a \emph{long} track and up to one \emph{short} track. The topology recognition selection criteria will eliminate those events in which one of the electrons has too low energy to produce a blob, around a 5\%. The remaining signal events rejected by the blob method are the ones for which the algorithm fails to reproduce the real physical track (\emph{track coverage} conditions explained in subsection \ref{subsec:Discrimination_blobs}), due to excessive straggling of the track. Finally, the fiducial cut will reject those events which statistically deposit part of their energy close to surfaces.

Electron diffusion along the drift makes the tracking
more complicated since it widens the final electron cloud and reduces the charge per pixel. Moreover,
it is an effect that depends on the distance of the event to the readout. A critical consequence of diffusion is that different tracks produced closely may merge and fake a two--electron--single--track topology. The blob identification is also hindered by the presence of diffusion. Although our algorithm is able to partially compensate these effects by properly adjusting its parameters (e.g. by means of the different pixel charge thresholds defined), the diffusion will affect the available topological information and the final discrimination capabilities. In an attempt to quantify this effect, a low diffusion scenario is studied in the next section.

\subsection{A low diffusion Xe mixture} \label{subsec:results_xenon_TMA}

In this subsection a low diffusion xenon mixture example is analyzed in order to study whether it could imply an improvement on the topological recognition of the events. A representative value of $\mathrm{\sim 100 \mu m/ \sqrt{cm}}$ for both transversal and longitudinal diffusion coefficients has been chosen in order to study the impact of this parameter in pattern recognition algorithms.

As in the pure xenon case, each selection criterion is applied to a population which has passed previous cuts. Table \ref{tab:rejfaCutsXeTMA} shows the relative background reduction factors (in \%) for different background sources, and table \ref{tab:efficiencyCutsXeTMA} the relative $\beta\beta0\nu$ signal efficiencies of each selection criterion.

\begin{table}
\caption{Percentage of the surviving events of the different sources of background simulated after the successive application of the selection criteria in the RoI for a 1\,cm--length pixelized detector in a low diffusion xenon mixture at 10\,bar. The percentage of events showed is with respect to the population passing previous cuts.
\label{tab:rejfaCutsXeTMA}}
\begin{indented}
\lineup
\item[]\begin{tabular}{@{}*{7}{l}}
\br 		                              &            &              &                 &\centre{3}{Fiducial method} \\
\ns
&&&& \crule{3} \\
 Origin & Isotope  & Track meth.        &Blob meth.           &Lateral    &Bottom        &Top\\
\mr
\lineup
   Lateral   & $^{208}$Tl     &  \07.1$\pm$0.3    &11.7$\pm$1.6     &83.1$\pm$16.1     &100.0$\pm$20.2     &\098.0$\pm$19.9 \\
  	vessel  & $^{214}$Bi      &  16.0$\pm$0.6   &12.6$\pm$1.2     &93.4$\pm$12.2     &\099.1$\pm$13.2      &100.0$\pm$13.4 \\[1mm]

  Field & $^{208}$Tl &  \04.4 $\pm$0.2    &14.7$\pm$1.6     &64.9$\pm$10.7     &100.0$\pm$18.1     &\096.7$\pm$17.7 \\
  	cage				& $^{214}$Bi	&  42.1$\pm$0.4   &10.6$\pm$0.2     &19.8$\pm$\01.1      &\099.3\0$\pm$6.9       &\098.5\0$\pm$6.9 \\[1mm]

 \multirow{2}{*}{Readout}& $^{208}$Tl&  \02.4$\pm$0.1     &23.9$\pm$2.2     &92.8$\pm$10.8     &\014.9\0$\pm$3.5      &100.0$\pm$30.9 \\
  						& $^{214}$Bi &  37.2 $\pm$0.5    &30.2$\pm$0.8     &89.8$\pm$\02.9    &\0\00.2\0$\pm$0.1       &100.0$\pm$81.6 \\[1mm]

 \multirow{2}{*}{Cathode} & $^{208}$Tl &  \04.8$\pm$0.1     &18.0$\pm$1.4     &93.6$\pm$\09.4     &100.0$\pm$10.2     &\024.1\0$\pm$4.0 \\
  						& $^{214}$Bi &  57.4$\pm$0.6    &26.9$\pm$0.5     &83.6$\pm$\01.9     &100.0\0$\pm$2.4      &\0\00.3\0$\pm$0.1 \\
\br
\end{tabular}
\end{indented}
\end{table}

A comparison between tables \ref{tab:rejfaCutsXeTMA} and \ref{tab:rejfaCutsXe} shows that the low diffusion hardly changes the rejection due to the track method since pixels with too little charge had been removed also in the case of a high diffusion. Only in the case of the readout contamination, somewhat improved factors are achieved due to the fact that low diffusion is especially relevant at low drift distances. However, it is in the topological recognition of blobs where almost an additional factor 3 of improvement in rejection is obtained, mainly due to the better recognition of secondary photons interactions near the main track: the improvement is more important in the case of events caused by photons (case of $^{208}$Tl contaminations) than in those which include also primary electrons (as those from beta spectrum of $^{214}$Bi on surfaces). The fiducial rejection it is slightly worse than in the case of pure Xe due to the fact that events are now less extended.

\begin{table}
\caption{Efficiency of the simulated $\beta\beta0\nu$ events after the successive application of the selection criteria. Connection is applied over the events detected in the RoI and then number of surviving events showed is calculated with respect the previous one. The events are simulated for a 1\,cm--length pixelized detector in a low diffusion xenon mixture at 10\,bar.
\label{tab:efficiencyCutsXeTMA}}
\begin{indented}
\lineup
\item[]\begin{tabular}{@{}*{7}{l}}
\br
  		                              &            &              &                 &\centre{3}{Eff. Fiducial vetoes} \\
\ns
&&&& \crule{3} \\
 Origin & Isotope  & Eff. Track        & Eff. Topology            &Lateral    &Bottom        &Top\\
\mr
\lineup
      Target &$^{136}$Xe                     &76.1 $\pm$0.4   &78.9 $\pm$0.5     &95.4$\pm$0.6					 &99.4$\pm$0.6           &98.9$\pm$0.6   \\
\br
\end{tabular}
\end{indented}
\end{table}

Regarding signal (table \ref{tab:efficiencyCutsXeTMA}), the use of a quencher to decrease the diffusion allows a better determination of two separated tracks and some small extra depositions are now distinguished apart from the main track. The track--line is also sharper (see figure \ref{figure9}) for both, signal and background events, and, therefore \emph{track coverage} parameter $R_{\rm cv}$ (table \ref{tab:blob para2}) needs to be chosen more stringent to reject background. For this reason, the efficiency after the track and blob method is lower than in the case of pure Xe, but it is increased after the lateral fiducial method since less events are close to the wall.

\subsection{Impact of Pattern recognition techniques on background levels}\label{subsec:evaluation}

As already mentioned, the aim of this work is not to determine final background levels, for which a detailed background model including final detector geometry and components with their corresponding contaminations would need to be built. However, it is interesting to address the issue of whether the performance of the algorithm developed could be generically sufficient to reach competitive background levels in the RoI that, as explained in the introduction, must be of the order of  $10^{-3}$\,c\,keV$^{-1}$\,kg$^{-1}$\,year$^{-1}$ or below. To answer this question we do the following exercise.

We have normalized the populations of simulated events in the different volumes of the simulated geometry with the following assumed volumetric contaminations: 1\,$\mu$Bq/kg for $^{208}$Tl and 10\,$\mu$Bq/kg for $^{214}$Bi for copper (vessel, field cage wires and cathode), and 10\,$\mu$Bq/kg for $^{208}$Tl and 5\,$\mu$Bq/kg for $^{214}$Bi for teflon (field cage body). These values are conservative upper limits of $^{208}$Tl and $^{214}$Bi contaminations typically obtained for copper and teflon samples in the literature \cite{Alvarez:2013, Aprile:2011}. The background level in the RoI derived from these numbers amounts to $(2.0\pm0.1)\times 10^{-4}$\,c\,keV$^{-1}$\,kg$^{-1}$\,year$^{-1}$  in the case of pure xenon, while it is considerably lower, $(0.9\pm0.1)\times 10^{-4}$\,c\,keV$^{-1}$\,kg$^{-1}$\,year$^{-1}$ for the low diffusion scenario. In both cases much below the specified $10^{-3}$\,c\,keV$^{-1}$\,kg$^{-1}$\,year$^{-1}$.

Most probably, in a real TPC the background will be dominated by other internal components linked to more specific detector features like adhesives, soldering, electronic components, or other elements of the sensors, readout, connectors or feedthroughs. Of these, the ones closer to the gas volume will be the most critical. $^{222}$Rn emanation, as explained before, may add up as an additional surface contamination of $^{214}$Bi.
Keeping the generality of our work, we have proceeded now reversely to calculate which level of surface contamination of the readout plane translates into a background level of $10^{-3}$\,c\,keV$^{-1}$\,kg$^{-1}$\,year$^{-1}$ after discrimination. The resulting values are of  $\sim 2.5$\,$\mu$Bq/cm$^2$ for $^{208}$Tl and $\sim 15$\,$\mu$Bq/cm$^2$ for $^{214}$Bi, for the case of pure xenon.  In the low diffusion scenario this contamination could be 3 times higher.
This calculation gives us an idea of the level of internal contamination (simplistically expressed as an internal surface contamination of the readout) tolerated in view of the discrimination capabilities of our algorithm. The obtained values, although stringent, are within the typical goals of screening campaigns carried out by current low background experiments.

To conclude, together with state of the art radiopurity, the topological discrimination of a Xe gas TPC as implemented in our algorithm is capable of bringing the levels of background of such a TPC to the required levels of the next generation experiment.

\section{Conclusions and outlook}

In this paper, we have done a detailed simulation of the topology of $\beta\beta$ $^{136}$Xe events, as well as of typical  background events in the range of interest of the $\beta\beta0\nu$ decay mode, in a high pressure Xe TPC with pixelised readout. A pattern recognition algorithm has been  developed and applied to the simulated topologies to discriminate signal and background, and determine  efficiency and rejection factors.

The background events considered come from decays of $^{208}$Tl or $^{214}$Bi in different internal components of the  detector. The simulation includes the particle transport and interaction with the detector (Geant4) as well as  the drift and diffusion of the ionization cloud in the TPC and the digitization of the signal in a pixelized  readout of 1$\times$1\,cm$^2$ pixel size. The simulations are carried out within a generic TPC geometry including about  $\sim$ 100\,kg of gas Xe at 10\,bar pressure.

The rationale of the discrimination criteria relies on the following features of signal events: the absence of  energy deposits close to the walls (fiducial cut), the presence of only one long track and the identification  of two larger energy depositions (blobs) at the two ends of the track. The presence of a second track or  secondary energy depositions are also dealt with by the algorithm. These basic ideas are implemented in a pattern recognition algorithm based on graph theory concepts.

The rejection factor for each of the background components studied has been determined, as well as the  breakdown of this factor into each of the independent criteria separately. In overall, the algorithm is able  to reduce the events falling in the energy RoI by about 3 orders of magnitude. The efficiency of the algorithm  on signal events is maintained at a 40\%.

Our result is comparable to the one previously achieved by visual inspection in \cite{{Wong:1993htw}} even with the relatively  large diffusion of pure Xe, and it is now implemented in a automated algorithm.

The use of our algorithm with a lower diffusion gas has been tested and an improvement of an extra factor 3 in  the rejection factor is easily achieved, while keeping the same signal efficiency. This improvement is mostly  concentrated on the criterion of identifying the two blobs. We consider this result conservative as it is  probably limited by the pixel size used, of 1$\times$1\,cm$^2$, not optimized for the low diffusion case. We plan now to  continue studying this case with smaller pixel sizes.

In general the nature of the events surviving the discrimination criteria has been briefly discussed to understand the limitation of  the algorithm. Extra energy deposition (by, e.g., x-ray emission or low energy photon radiation) close to the main track or  excessive straggling of the electron track are sometimes cause of bad reconstruction of the events. Several  ideas are put forward to correctly reconstruct those events, and there are prospects to further improvement of the algorithm. Future work will be focused on a deeper study of misidentified events to achieve a better blob identification. Positive results may open the possibility of increasing the gas pressure, resulting in a more compact design, less electronic channels or the ability of working with larger volume detectors.

\section*{Acknowledgements}
We are grateful to our colleagues of the groups of the University of Zaragoza, CEA/Saclay and
our colleagues from the NEXT and RD-51 collaborations for helpful discussions and encouragement. We especially thank D. Gonz\'alez-D\'iaz for careful reading of the manuscript and giving useful comments. We acknowledge support from the European Commission under the European Research
Council T-REX Starting Grant ref. ERC-2009-StG-240054 of the IDEAS program of the 7th EU
Framework Program. We also acknowledge support from the Spanish Ministry of Economy and
Competitiveness (MINECO), under contracts ref. FPA2008-03456 and FPA2011-24058, as well as
under the CUP project ref. CSD2008-00037 and the CPAN project ref. CSD2007- 00042 from the
Consolider-Ingenio 2010 program of the MICINN. Part of these grants are funded by the European
Regional Development Fund (ERDF/FEDER). F.I. acknowledges the support from the Eurotalents
program and D.C.H. of the Univ. Zaragoza, under the program PIF-UZ-2009-CIE-03.

\end{document}